\documentclass[letterpaper,twocolumn,10pt]{article}
\usepackage{usenix2019_v3}

\usepackage{graphicx}
\usepackage{booktabs}
\usepackage{array}
\usepackage{amsmath}
\usepackage{enumitem}
\usepackage{float}
\usepackage{subcaption}
\usepackage{tikz}

\makeatletter
\g@addto@macro{\UrlBreaks}{\UrlOrds}
\makeatother

\begin{document}

\raggedbottom

\date{}

\title{\Large \bf HiLO: Towards Product-Level Graceful Degradation with\\
a Quality-Aware, Load-Aware Control Framework}

\author{
{\rm Knut Magne Risvik}\\
Microsoft
\and
{\rm Zhuowei Li}\\
Microsoft
\and
{\rm Nirupam Purushothama}\\
Microsoft
\and
{\rm Hunter Su}\\
Microsoft
\and
{\rm Greg Dunham}\\
Microsoft
\and
{\rm Yan Reznikov}\\
Microsoft
}

\maketitle
\pagestyle{plain}

\begin{abstract}
Graceful degradation is a key mechanism employed by large-scale systems to handle overload conditions. Its importance has grown further as GPU-intensive LLM inference introduces persistent capacity constraints that cannot be addressed through provisioning alone. Common degradation approaches rely on service-level mechanisms such as load shedding and circuit breakers, or on feature-level switches (some manually controlled) to preserve availability and quality. While effective at protecting services, such approaches lead to untestable product states and thereby unpredictable product quality.

We present HiLO (High Load Optimizer), an automated product-level degradation system designed to handle graceful degradation at scale. HiLO maps observed load to a product state lattice: a small set of pre-validated product states, partially ordered by degradation severity. Dependencies are resolved at design time enabling deterministic and continuous control without requiring runtime dependency discovery. The same abstraction applies unchanged across heterogeneous services, including GPU-intensive deep models, LLMs and their serving layers.

At Microsoft, HiLO is deployed in Bing (serving over a billion monthly active users), where it manages on average two traffic surges per week across all datacenters, bounding OneDCG (relevance) drop to under 0.5 points (the operational target) even under sustained load exceeding 2$\times$ baseline peak. HiLO complements existing service-level protections with product-level coordination and makes product quality measurable, testable and predictable.

\end{abstract}

\section{Introduction}
\label{sec:intro}

Web products such as Bing serve over a billion monthly active users~\cite{stats:bing}. Online traffic has grown significantly over the past several years~\cite{feldmann:lockdown, cisco:air, cloudflare:radar2025}. In parallel, AI-integrated search and chat systems have seen rapid growth in adoption, contributing additional demand~\cite{alvi:bing140m}. Additionally, web services face increased volumes of DDoS attacks, automated bot traffic and large-scale scraping activity~\cite{cloudflare:radar2025, imperva:badbot2025}.

This sustained growth in traffic, coupled with transient capacity losses from machine failures, new feature deployments, and faulty services keep product availability at risk during overload. Moreover, search engines and similar products increasingly leverage LLMs~\cite{xiong:searchllm}, creating significant strain on GPU capacity. This requires all services dependent on them to frequently cope with increased latencies and bottlenecks. These fluctuations routinely occur on time scales faster than human intervention can adapt~\cite{dean:tail, suresh:c3}, necessitating automated policies that react quickly to prevent overload conditions from spiraling into outages.

To mitigate this risk, most online services rely on two broad strategies: overprovisioning and service degradation. Overprovisioning maintains significant headroom to absorb unexpected traffic increases but is expensive and inefficient at scale. In practice, capacity constraints (particularly for specialized resources such as GPUs) are persistent~\cite{cnbc:gpushortage, aws:gpuchallenges}. Moreover, the scale and timing of demand spikes remain difficult to predict~\cite{cloudflare:ddosmodels}, thereby reducing the efficiency of overprovisioning.

Hence, many systems leverage graceful degradation policies to ensure their survival under overload conditions~\cite{defcon, netflix:degradation, uber:degradation}. Especially in Bing's case, additional headroom is infeasible, making graceful degradation essential. Commonly, service-level protections such as load shedding, throttling, and circuit breakers~\cite{sre:book, hystrix} are employed to preserve availability. While effective at protecting individual services, these mechanisms cannot promise a baseline product quality because they operate locally and independently. This is because, in products composed of many interdependent services, independent degradation decisions interact in unexpected ways, causing sharp shifts in latency and quality that can far outlast the triggering event, a metastable dynamic~\cite{bronson:metastable, huang:metastable, huang:gray, mogul:nines} (e.g., capacity challenged services causing major dips in quality and service health; \S\ref{sec:pre-hilo}). Similarly, approaches that rely on manual knobs introduce a large interaction space across features and services, requiring continual maintenance~\cite{defcon}.

In this context, we introduce HiLO, an automated, quality-aware and load-aware product-level degradation system. Each service reports a load-factor, a normalized measure of its current load. HiLO maps the aggregate load-factor to a product state lattice: a small, finite set of pre-validated product states, partially ordered by degradation severity. Each state deterministically maps to coordinated service levels (each service's degradation level) across all participating services. States are created through a formal construction process by addressing dependencies at design time, enabling continuous and autonomous control without runtime dependency discovery. This abstraction applies unchanged across heterogeneous services, including GPU-intensive models and LLMs. Drawing on over six years of operating graceful degradation across Bing's datacenters, with HiLO in production since 2025, we share the design insights and operational lessons, and evaluate HiLO across multiple production scenarios.

\section{Background and Motivation}
\label{sec:background}

As described in the introduction, demand capacity mismatch is a persistent challenge. Here we discuss past learnings, the metrics used to evaluate the implications of graceful degradation, and design gaps.

\subsection{Five Metrics to Watch}
\label{sec:metrics}

The following metrics are important to evaluate the implications of a graceful degradation policy.

\begin{description}[leftmargin=0pt]
\item[Availability:] This metric captures whether the graceful degradation policy preserves basic functionality and prevents service outages~\cite{sre:book}.

\item[Quality:] This is a combination of several signals. The primary one at Bing, OneDCG is a normalized variant of discounted cumulative gain~\cite{jarvelin:dcg}, used to measure relevance of results. In addition, error signals such as no-result rates and user-facing experiment metrics such as content freshness and engagement (e.g., reformulation rates) are measured to validate the quality under each product state (\S\ref{sec:eval:experiments}).

\item[Stability:] This metric determines the percentage of queries that receive consistent responses across repeated executions under identical conditions, capturing the predictability of product behavior:
\[
\text{Stability}(\%) = \frac{|\{(q_i, q_j) : r_i = r_j\}|}{|Q_{\text{pairs}}|} \times 100
\]
where $Q_{\text{pairs}}$ is the set of query pairs $(q_i, q_j)$ issued under identical observable context (same query, same time window), and $r_i$, $r_j$ are their respective responses. A pair is counted as stable if the responses are identical. Higher instability typically indicates underlying systemic issues (i.e., abnormal service behavior) and hence is closely monitored.

\item[Recovery velocity:] This indicates the time taken for stability and quality to return to pre-incident levels after an overload event, and hence the time it takes for underlying services to return to normalcy.

\item[Capacity savings:] This metric quantifies the amount of capacity reclaimed during the graceful degradation time. We use ``load-factor'' (LF)~(\S\ref{sec:architecture}) to measure the capacity utilization level of a service during runtime.

\end{description}

\subsection{Evolution from Basic Graceful Degradation to HiLO - Quality Awareness}
\label{sec:pre-hilo}

At Bing, for many years, the degradation policy focused on ensuring high availability by protecting individual services during overload conditions. Services performed two primary activities: (1)~\textit{Monitor health:} Individual services self-monitored local load or health signals such as CPU utilization, request queue depth, latency, or error rates. (2)~\textit{Adjust service level:} Based on the current load situation, services independently applied predefined degradation actions when thresholds were crossed. Common mechanisms included disabling non-critical features, reducing computation fidelity, limiting request fanout, increasing cache reuse, or shedding low priority traffic.

This allowed Bing to maintain availability well above 99.95\%. However, this policy introduced three fundamental challenges. (1)~The current product state became difficult to reason about, (2)~it became difficult to provide consistent baseline quality guarantees to the users and (3)~inter-service dependencies led to localized orchestration and control loops. Some components attempted to directly control their immediate dependencies using feedback-based mechanisms (e.g., PID-style control~\cite{astrom:pid}), which in practice complicated testing, debuggability, measurability and experimentation at the product level. The following incident illustrates how a local degradation policy fails to protect overall system health and cannot hold baseline quality. 

\begin{figure}[H]
\centering
\includegraphics[width=\columnwidth]{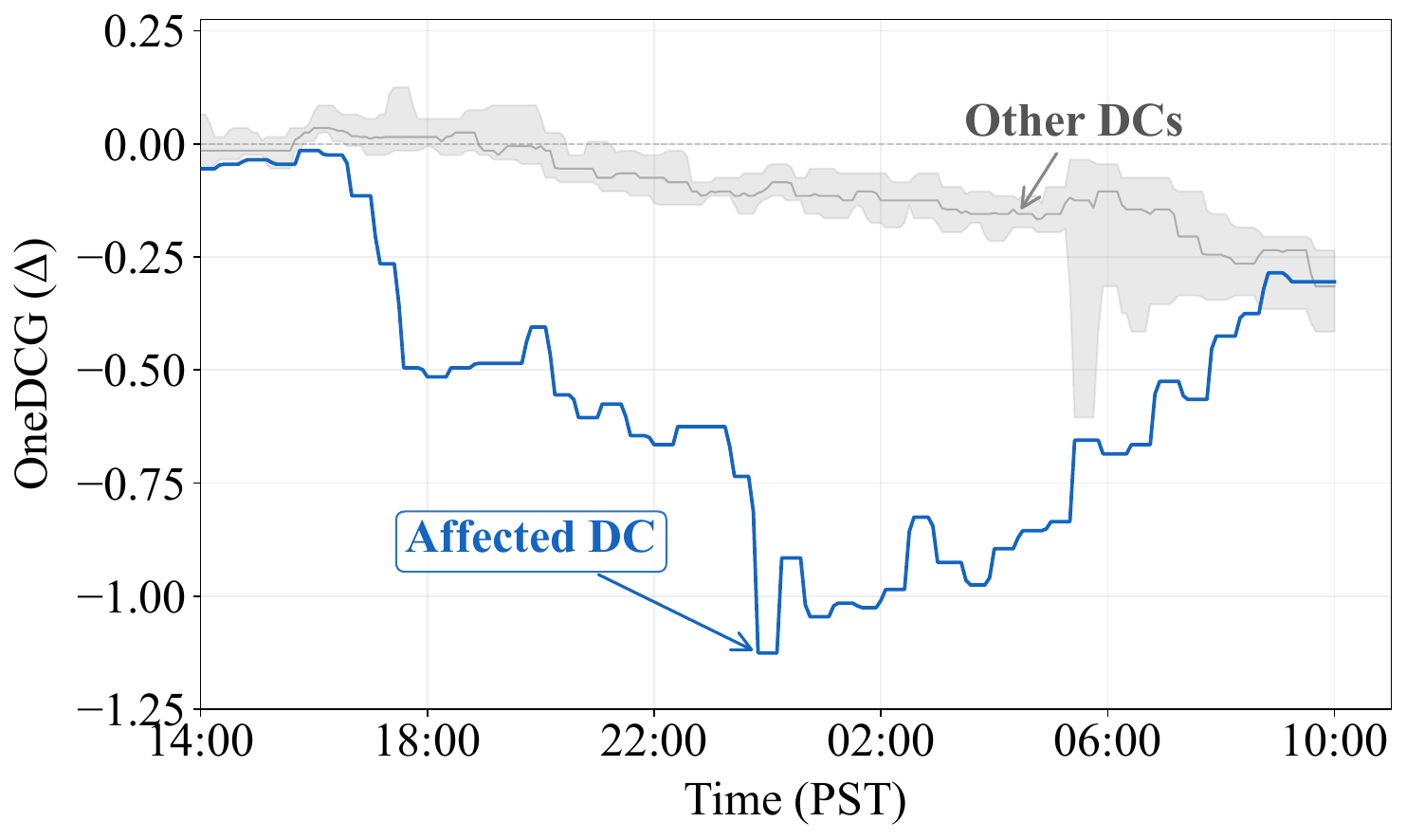}
\caption{Per-datacenter OneDCG (relevance) on the day of incident. Overload in one datacenter causes OneDCG to drop \textasciitilde1.11 points. No coordinated mechanism existed to handle this gracefully.}
\label{fig:incident-quality-prehilo}
\end{figure}

Figure~\ref{fig:incident-quality-prehilo} depicts the movement in result quality after a feature deployment increased the call volume to a model-serving service shared by multiple workflows. The change passed flight testing at fractional traffic, where capacity was sufficient. After full deployment, one datacenter was already running near its capacity limits when a transient load spike (under 10 minutes) pushed it past its capacity limit. Following its own service-local policy, the already pressured shared model-serving service degraded to protect itself and stayed available. But its callers kept calling under their own policies with no coordinated way to reduce demand, so that the bottleneck surfaced upstream as elevated latencies and cascading errors, causing a OneDCG decline of 1.11 points; the service survived, but quality suffered significantly.

\begin{figure}[H]
\centering
\includegraphics[width=\columnwidth]{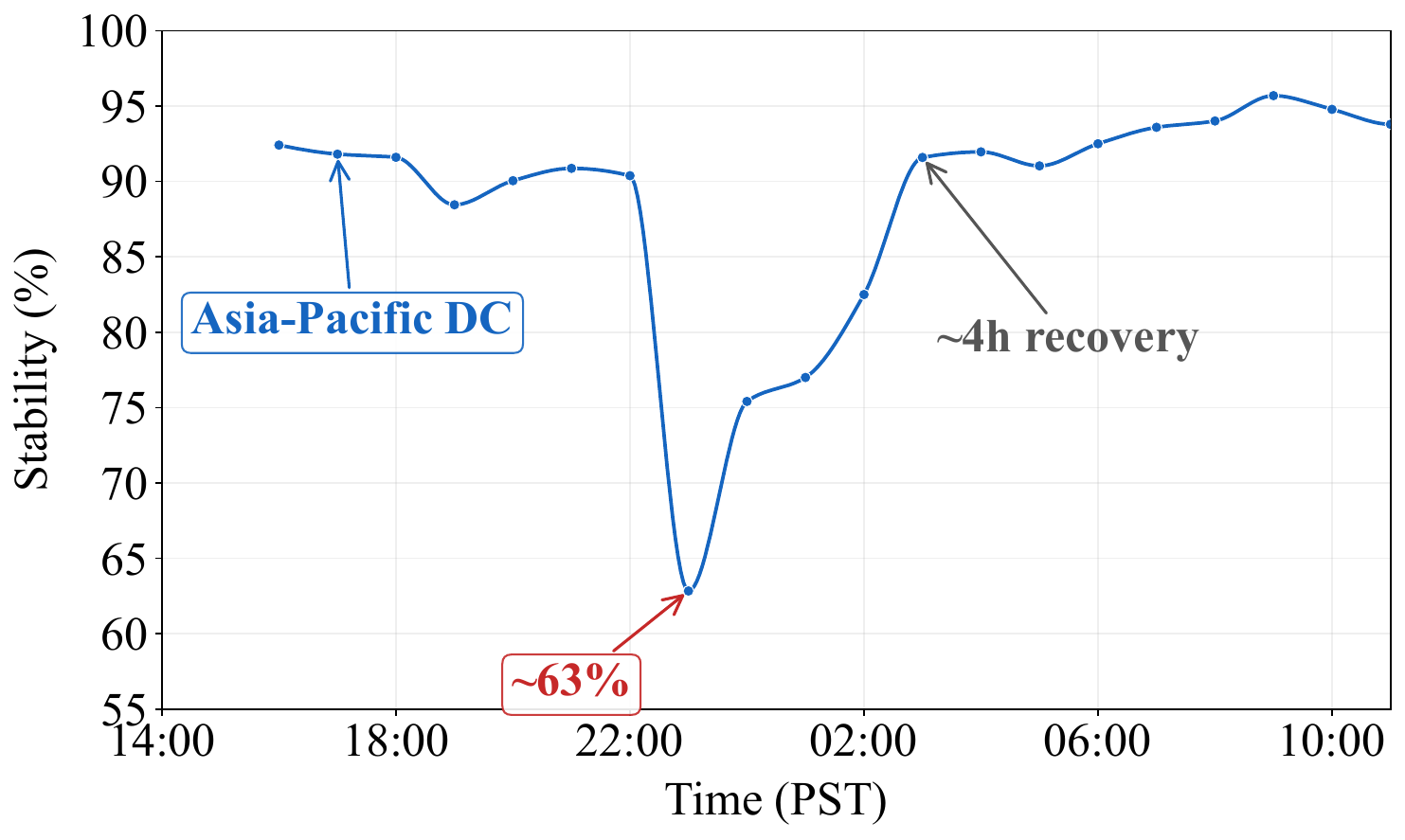}
\caption{Product stability in the affected DC on the day of incident. A transient load spike (under 10 minutes) causes system-wide instability. Stability drops from \textasciitilde90\% to \textasciitilde63\% and takes over four hours to recover.}
\label{fig:incident-stability-prehilo}
\end{figure}

Figure~\ref{fig:incident-stability-prehilo}, related to the same incident, depicts how the product stability fell from \textasciitilde90\% to \textasciitilde63\% and took over four hours to recover (indicating prolonged abnormality with underlying services even after the spike receded). This incident triggered a severity-2 escalation. OneDCG fluctuated for another two days (whenever load increased) and recovered to its pre-incident baseline after a fix to reduce the additional dependency calls (by the initial deployment) was deployed.

Performance isolation guarantees for individual services always existed at Bing (each service owns its capacity, so one service's overload cannot starve another), and service-local degradation (at least throttling) was the norm. This particular incident highlights the risk of an uncoordinated degradation policy, which could result in metastable failures~\cite{bronson:metastable} and prolonged quality loss, even when underlying service availability is maintained. This dynamic is not specific to model serving. It can arise at any shared, capacity-constrained service. At Bing this is an increasingly common issue with models and their serving layers given persistent GPU capacity constraints.

The investigation exposed several structural gaps. Services were allowed to operate near their capacity limits, relying on local protection to handle overflows. When the system reached its limit, no mechanism existed to reduce demand on the shared service before the spike or to coordinate recovery afterward, and debugging was hard because the product-level state was unclear. Ultimately, quality (OneDCG) during this event was an emergent behavior rather than a controlled outcome. This calls for a fast-acting, product-level policy that acts before overload by formalizing load signals, dependencies, and their coordination, which is what HiLO provides.

\subsection{Related Work}
\label{sec:related}

For graceful degradation many systems either adopt feature-centric degradation or admission control mechanisms or a combination of techniques to preserve service availability. Feature-centric control is achieved by degrading optional product functionality in order to reclaim resources while preserving core behavior. A canonical work is Brownout~\cite{klein:brownout}. It suggests splitting an application into components where optional components can be isolated and degraded. A dimmer at application level indicates how frequently these optional components are deactivated and a controller controls the dimmer based on measured latency.

Meta's DefCon system~\cite{defcon} is another representative example of feature-level degradation at scale. It exposes binary degradation knobs that allow operators to selectively and manually control feature execution under overload. It assumes strong isolation between feature execution paths, noting that features at Meta are typically implemented as separate RPCs with minimal control flow interaction. It achieves control through config-driven operations by allowing operators to utilize a relational database to determine the outcome of choosing a set of control knobs. Designed for infrequent site emergencies with human driven operation, Meta also notes the operational cost of maintaining hundreds of knobs that grow in number over time.

Google's SRE guidance~\cite{sre:book} discusses overload handling around load shedding (per-task throttling based on CPU, memory, queue length or request priorities) and graceful degradation. It emphasizes the importance of testing degraded modes regularly and maintaining operational visibility through monitoring and alerts. SEDA~\cite{seda} is another canonical line of work that structures applications into explicit event-driven stages and queues preventing resources from getting overloaded. Use of dynamic resource controllers is suggested at each stage to keep it within control limits. The controller observes performance and load, automatically adjusting the stage's resource usage. Its extension presents an adaptive admission control~\cite{welsh:overload} mechanism that attempts to meet the P90 response time target by filtering requests at each stage. Instead of CPU utilization and fraction of users, this work focuses on response time as the primary metric. Based on the calculated P90 latency at this moment, the controller adjusts the admission rate.

Dagor~\cite{zhou:dagor} from WeChat decouples overload control from service logic, making it service agnostic. Each server monitors the average waiting time of requests in its pending queue (window based on time and request count) to detect overload. On detection, it performs admission control using business priority (e.g., login and payment requests are higher priority than instant messaging). Dagor introduces user priority to provide finer-grained shedding within each business priority tier. Business and user priorities are composed into a compound admission level that is adjusted adaptively. Overloaded servers also piggyback their admission thresholds to upstream callers, enabling collaborative early shedding. Breakwater~\cite{breakwater} targets overload control for microsecond scale RPCs, where the cost of dropping a request is comparable to processing it. It presents a server driven admission control scheme that prevents overload through the use of server issued credits. Here, a credit represents availability at the server to process a single request by the client that receives the credit. Server load is effectively managed through the distribution of total credits and the rate of distribution. Rajomon~\cite{rajomon} extends this line of work with market-based approach where services set prices for their APIs and clients attach tokens to requests, enabling decentralized and coordinated rate limiting across the call graph.

In practice, comparable strategies often combined with circuit breakers~\cite{hystrix, resilience4j} are reported across industry. Sometimes feature degradation is achieved through admission control. Netflix describes selectively throttling lower priority requests so that playback experience remains uninterrupted ~\cite{netflix:degradation, netflix:servicelevel}. It describes the need to continuously prioritize requests (including finer-grained prioritization at service layer), progressively throttle based on priority and validate through chaos testing. Similarly, Uber's QALM framework sheds non-critical requests based on criticality tiers during overload. It creates separate queues for endpoints each with its own overload detector (that monitors latency). Then using request criticality drops requests only in the affected queue thereby isolating the impact and preserving core trip and infra requests~\cite{uber:degradation}. AWS~\cite{amazon:primeday} documents graceful degradation as a reliability pattern, recommending that services transform hard dependencies into soft ones to preserve availability under failure.

When compared to the above methods, HiLO targets systems where services are tightly interdependent, capacity is heterogeneous, and overload requires continuous autonomous management rather than manual intervention or binary feature toggling. For example, large search and LLM-serving stacks such as Bing serve queries that span across multiple CPU-bound services and GPU-bound ranking/LLM models with complex dependencies. The design is detailed in \S\ref{sec:hilo}.


\begin{figure*}[t]
\centering
\includegraphics[width=0.75\textwidth]{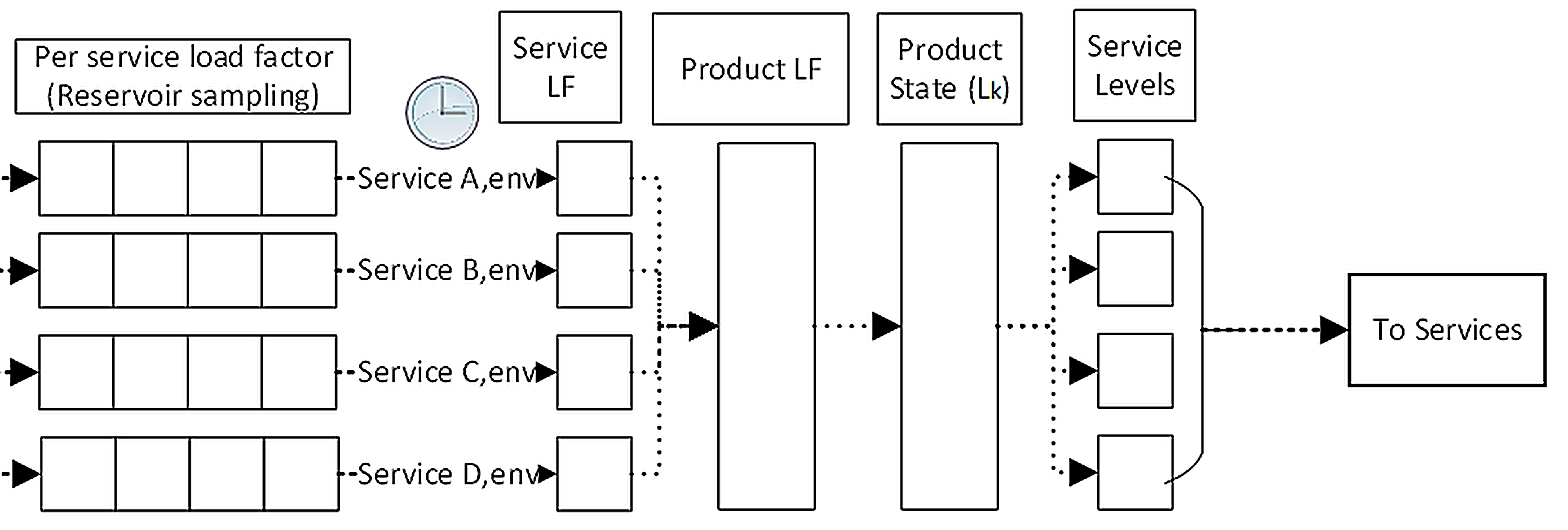}
\caption{HiLO internal decision pipeline. Services periodically report load signals, which are sampled and aggregated in HiLO. This is used to determine the load at product level and the corresponding product state ($L_k$), which determines the service levels for participating services. HiLO disseminates this back to the services which enforce them.}
\label{fig:hilo-internals}
\end{figure*}

\section{HiLO (High Load Optimizer)}
\label{sec:hilo}

Guided by the gaps identified in \S\ref{sec:pre-hilo}, we now describe HiLO's core abstractions and architecture.

\begin{figure}[t]
\centering
\includegraphics[width=\columnwidth]{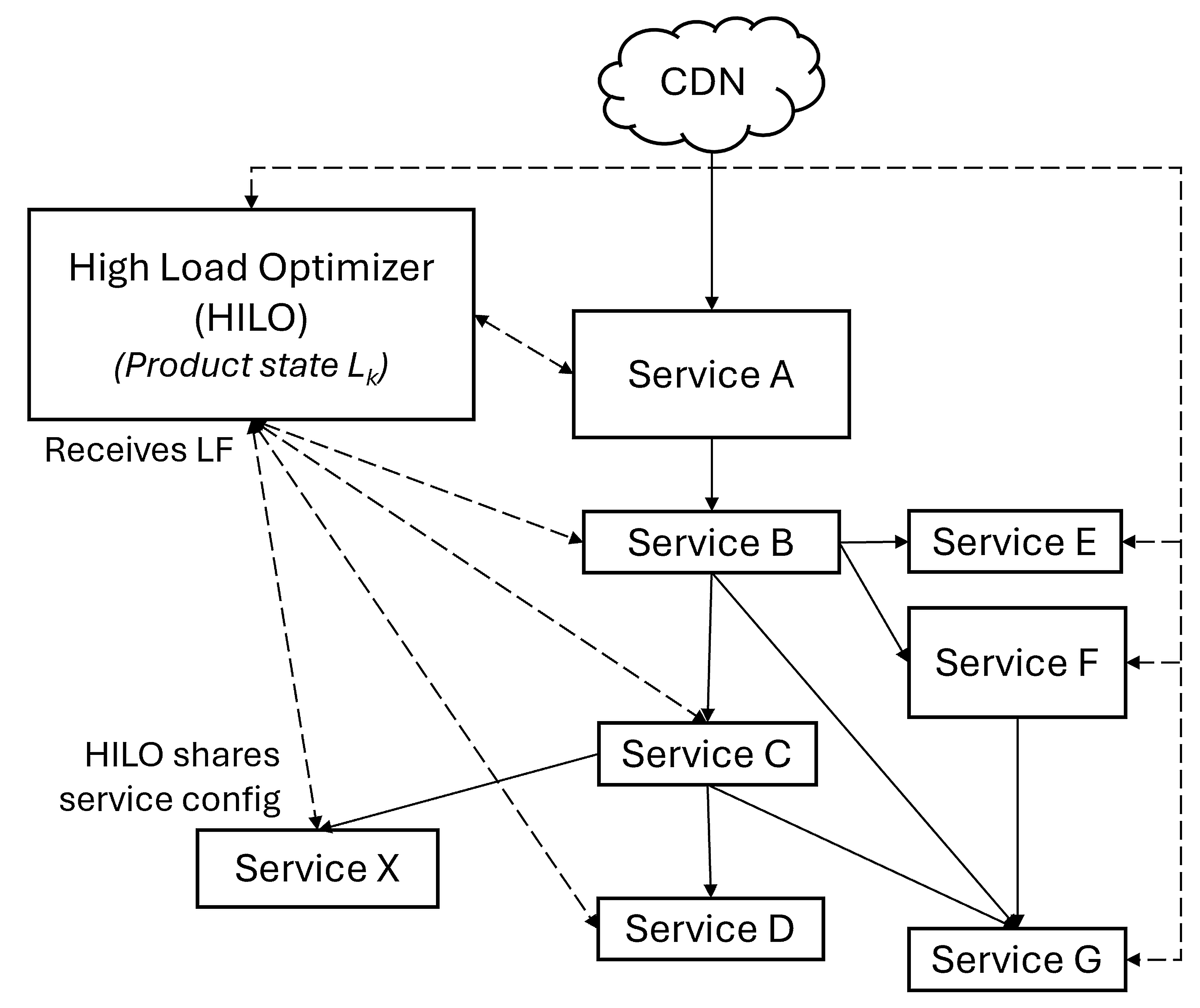}
\caption{System overview showcasing HiLO's position in the stack. HiLO operates out of the request path, aggregates load signals reported by individual services, selects a product state ($L_k$), and disseminates coordinated configuration decisions to participating services.}
\label{fig:system-overview}
\end{figure}

\begin{figure}[t]
\centering
\includegraphics[width=\columnwidth]{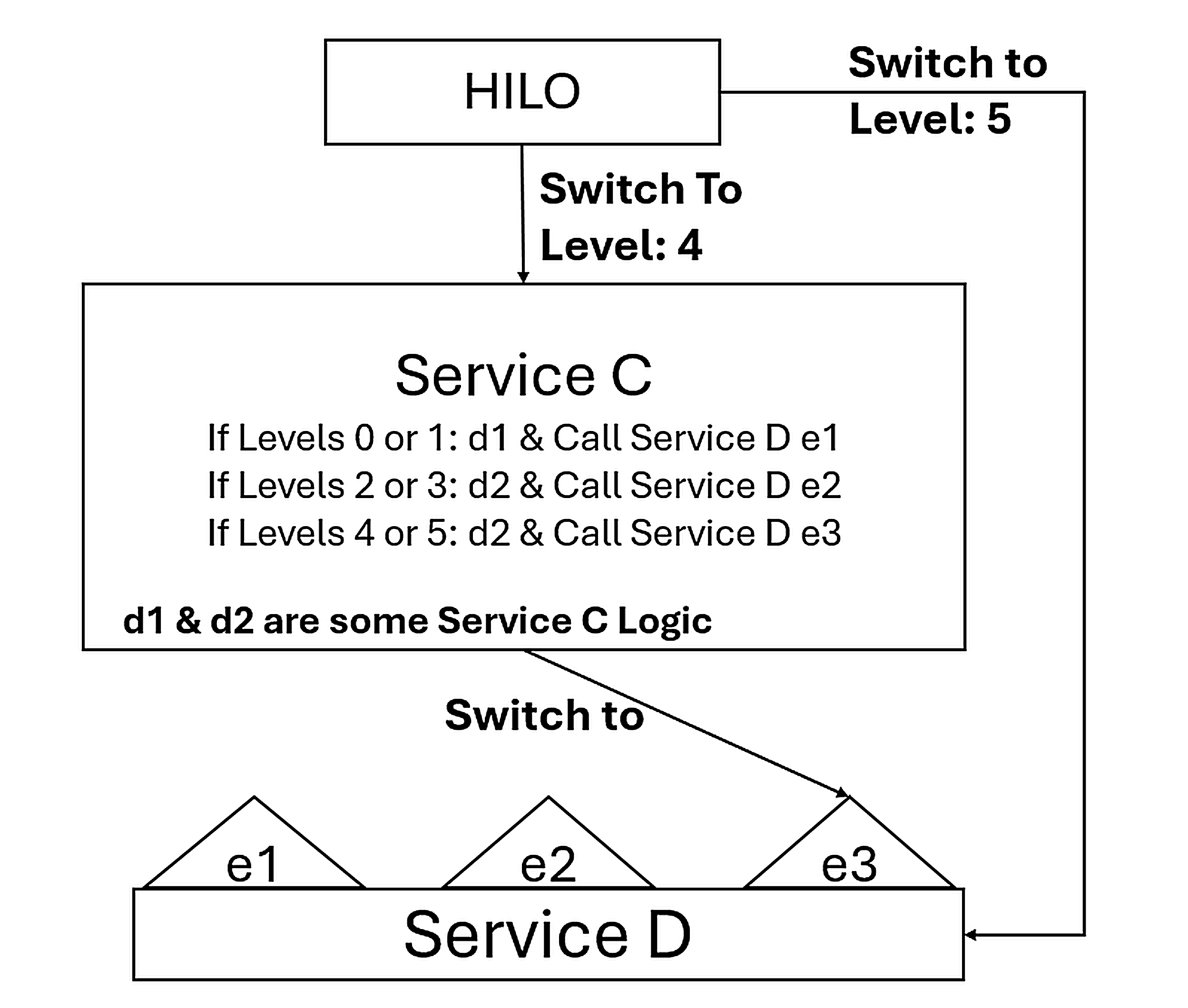}
\caption{Dependency handling under coordination at product level. HiLO determines service levels based on the active product state. This determines predefined execution logic and downstream service variants, enforcing consistent dependency resolution across services.}
\label{fig:dependency-handling}
\end{figure}

\subsection{Load-Factor and Product State Lattice}
\label{sec:architecture}

HiLO is built on two core abstractions: a \textit{Load-factor} that summarizes service-level capacity pressure, and a \textit{product state lattice} that defines the space of valid, coordinated degradation behaviors.

Figure~\ref{fig:hilo-internals} shows HiLO's internal pipeline. Each service emits a Load-Factor (LF), in real time, a normalized scalar signal reflecting its current service load. LFs are derived locally from service-specific indicators such as CPU or GPU utilization, request queue depth, tail latency, or dependency backpressure. How each service chooses this signal, and adapts it as its workload evolves, is discussed in \S\ref{sec:limitations}. HiLO operates outside the critical path (Figure~\ref{fig:system-overview}), requires no global capacity model or comparable units across services; it assumes only monotonicity (higher LF values indicate higher overload risk) and treats aggregated LFs as ordinal congestion signals rather than precise optimization inputs. In Bing, LFs are reported on a bounded scale and aggregated per service using robust sampling (e.g., reservoir sampling~\cite{vitter:reservoir}) to dampen noise and short-lived spikes.

A product state represents a coordinated operating mode of the product under overload. Each product state deterministically maps to a predefined service level for every participating service (e.g., disabled features, simplified execution paths, reduced fan-out, or lower cost model variants). Product states are partially ordered by degradation severity, forming a lattice in which higher states dominate lower ones in terms of capacity reclamation while incurring bounded, pre-validated quality loss. In Bing, we use 6 product states per scenario (\S\ref{sec:further-controls}), with each service implementing 2-6 service levels depending on its flexibility.

The product state lattice is constructed at design time. Service owners define candidate service levels, quantify their capacity impact, and validate quality effects locally. Dependency effects are resolved during composition. Figure~\ref{fig:dependency-handling} illustrates this: a product state switch moves individual services to new levels (indicated by arrows), changing not only the service's own behavior but also how it invokes its callees. For example, both how often a caller calls a callee and which lower-cost variant it selects are independently controlled. Product states are finalized after validating them using end-to-end test and experimentation hooks (\S\ref{sec:observability}). At runtime, HiLO selects a single active product state per scenario, transitions monotonically under increasing load, and applies controlled recovery to avoid unnecessary oscillations. No per-request decisions, runtime dependency discovery, or dynamic call-graph inference are performed. In practice we label these states as L$_0$ through L$_5$, in order of increasing degradation severity.

\textbf{Positioning:} Unlike DefCon~\cite{defcon} which exposes a large number of feature-level knobs whose interactions determine product behavior implicitly, and Brownout~\cite{klein:brownout} which controls a single application without cross-service coordination, the lattice constrains online behavior to a small set of validated, coordinated product states. Unlike admission control and circuit breakers, which act locally to protect service SLOs, the lattice provides an explicit product-level quality baseline under overload. HiLO deliberately trades fine-grained optimality for predictability, testability, and operational clarity, which we find essential in large, dependency-rich systems combining services with GPU-constrained model and LLM pipelines.

\subsection{Deep Model \& LLM Specific Strategies}
\label{sec:ai-models}

As discussed in \S\ref{sec:pre-hilo}, applying HiLO to GPU-intensive deep models and their serving layers is critical for Bing due to capacity constraints. An alternative approach is to use per-request classifiers to route queries between expensive and cheap model variants at runtime. HiLO avoids this complexity. The expensive model is always used under normal conditions, and degradation to lower-cost variants is applied uniformly only when overload is detected. This ensures best OneDCG by default, at no additional quality loss (unlike classifiers). Table~\ref{tab:ai-strategies} summarizes the methods by which we bring heavily used models or their model-serving layers (at Bing) into HiLO's fold. The specific application of these strategies is discussed in the case studies section (\S\ref{sec:eval:g-casestudies}).

To illustrate how these strategies work together, consider a production ranking model whose degradation is coordinated across two services. At all degradation levels, capacity is reclaimed by progressively reducing multi-queries (MQ) and documents processed. At deeper levels, model layer reduction is additionally applied (see \S\ref{sec:eval:deep-models} for details). Less disruptive upstream actions are applied first, with model-level actions added at deeper levels. The sequence is determined and validated at design time. No additional architectural changes are required.

\begin{table}[H]
\centering
\caption{Strategies utilized by Bing to balance degradation and OneDCG for GPU constrained models, LLMs and their serving layers.}
\label{tab:ai-strategies}
\small
\begin{tabular}{@{}c>{\raggedright\arraybackslash}p{4.5cm}>{\raggedright\arraybackslash}p{2.8cm}@{}}
\toprule
\textbf{\#} & \textbf{Strategy} & \textbf{Severity} \\
\midrule
a. & Reduce multi-queries (MQ) & Mild to Moderate \\
b. & Reduce \# docs evaluated & Mild to Moderate \\
c. & Model layer reduction & Moderate to High \\
d. & Skip model & High \\
e. & Increase cache expiration & Mild to High \\
\bottomrule
\end{tabular}
\end{table}

\subsection{Product and Service-Level Construction}
\label{sec:product-levels}

This section details the three phase construction process outlined in \S\ref{sec:architecture}. Table~\ref{tab:product-levels} is a representative example of the product-state config. Each state represents a coordinated service-level combination. We present case studies of how levels and targets are realized for deep models and services (\S\ref{sec:eval:deep-models}) and an LLM product (Copilot Search, \S\ref{sec:eval:copilotsearch}).

\begin{table}[H]
\centering
\caption{Product state lattice: Product state and service-level mapping. Higher number indicates greater degradation.}
\label{tab:product-levels}
\small
\begin{tabular}{@{}lccc@{}}
\toprule
\textbf{Product State} & \textbf{Service-1 Level} & \textbf{\dots} & \textbf{Service-X Level} \\
\midrule
L$_0$ & 0 & \dots & 0 \\
L$_1$ & 1 & \dots & 0 \\
L$_2$ & 2 & \dots & 1 \\
L$_3$ & 3 & \dots & 1 \\
L$_4$ & 3 & \dots & 2 \\
L$_5$ & 4 & \dots & 2 \\
\bottomrule
\end{tabular}
\end{table}

\textbf{Capacity reclamation targets:} Each service's per-level reclamation targets are derived from a stack-wide survival goal. For example, if a product sets a target of 2$\times$ baseline-peak requests per second (RPS) survivability without throttling, each service then determines the number of levels required to support that goal and estimates the target load-factor reduction for each level at constant RPS. Headroom creation strategies may include reducing local compute, reducing demand from upstream services, or both. When a service is shared across products, the reclamation targets per product/scenario (\S\ref{sec:further-controls}) are determined independently. Runtime product-state selection, however, is relative. For example, a scenario's degradation can be driven by other scenarios' load on the shared service.

After that, service levels and product-state config are determined in three phases.

\textbf{Phase 1 - Service local degradation validation:} Service owners first identify a set of local tasks that could be degraded and estimate the impact. For each task they estimate the load-factor improvement and quality drop based on the current load-factor utilization, its trigger rate (i.e., execution frequency), the proposed alternative action, and its relative priority. If needed, these estimates are discussed with product teams to finalize the candidates. Then these actions are grouped together to a level based on their relative-priority and estimated overall service load-factor improvement. Service teams then conduct tests and experiments (A/B that are service local) to validate the estimates and then proceed to finalize the service levels. For example, this pertains to model layer reduction (\S\ref{sec:ai-models}) where a model's inference cost can be reduced at the model service itself.

\textbf{Phase 2 - Dependency resolution:} In this phase, callees that require upstream changes to reduce load coordinate with their callers to integrate cross-service degradation actions (e.g., those listed in Table~\ref{tab:ai-strategies}) into the caller's service levels. If this introduces new sub-levels, they are incorporated into the caller's level definitions. The same estimation, testing and experimentation steps from Phase~1 are repeated jointly by callee and caller to validate levels against metric movement, priorities and reclamation targets.

\textbf{Phase 3 - Product-states creation:} In this phase, product-state config is finally created in HiLO using the service levels created in the previous two phases. Product (Scenario - \S\ref{sec:further-controls}) owners now use the existing set of service levels and frame them into product states where every higher product state indicates further degradation in service levels. These candidate states are then subjected to end-to-end testing and experimentation (using end-to-end test-hooks and experimentation-hooks, described in \S\ref{sec:observability}). Candidate states that either cause unintended experiences or degrade user-metrics beyond targets are discarded or adjusted as needed. Finally, end-to-end capacity tests help understand if the desired effect (survival, quality maintenance and rapid stability recovery) is obtained when the entire system is subject to the peak survival goal.

\subsection{Further Controls in HiLO}
\label{sec:further-controls}

Production deployment revealed the need for additional controls beyond product-state selection. Distinct product verticals demanded independent degradation policies, and business stakeholders requested value-aware degradation. HiLO incorporates both without altering its core abstraction.

\textbf{Route Context (Scenario Identifier):} A scenario identifier represents a distinct operating context in which a product state applies, such as different request routes or entry points that share a common call graph (e.g., product verticals such as Web Search, Image Search, Video Search etc.). This allows multiple routes to operate under different product states concurrently, even when backed by the same set of services. HiLO separates policies between scenarios and relative priorities between routes. This enables coordinated but differentiated degradation behavior across different contexts.

\textbf{Request Priority:} Request priority is optional metadata that classifies traffic within a given route into priority buckets (e.g., high, medium, low). This enables finer-grained degradation by allowing lower-priority traffic to be degraded before higher-value requests~\cite{netflix:servicelevel}. HiLO publishes its policy to the services by request-priority. This ensures that degradation is enforced gradually without impacting valuable traffic first. Priority assignment (on the request path) is external to HiLO and propagated through the request path as metadata.

The frontend service (Service~A in Figure~\ref{fig:system-overview}) is responsible for identifying the route context and assigning request priority, ensuring that downstream services operate on consistent, end-to-end metadata. By incorporating these dimensions, HiLO enables controlled, quality-aware degradation across contexts.

\subsection{Monitoring, Testing \& Experimentation}
\label{sec:observability}

\textbf{Monitoring:} HiLO and rest of the services are seamlessly integrated into the logging and monitoring framework of the platform, requiring no additional architectural changes. For example, in Bing, on the server side, all impressions including headers and response details are logged at individual services and at Frontend. A dedicated logging pipeline stitches these logs and directs them to Magnetar (MT), an internal large-scale offline data processing platform built on Kafka~\cite{kreps:kafka}, Spark~\cite{zaharia:spark} and HDFS~\cite{shvachko:hdfs} with Titan as interactive visualization (Microsoft internal) and querying layer. For real-time reporting and operational monitoring, services upload metrics and counters to Geneva (similar to Azure monitor), which provides near-real-time operational visibility. This infrastructure is simply extended by logging three details at HiLO: the aggregated load-factors from participating services, the selected product state, and the corresponding service levels disseminated to each service. Services log the HiLO-assigned state alongside each impression. Existing dashboards consume this additional metadata to report degradation behavior at product level as needed.

\textbf{Test hooks:} Before deploying new product states or service levels, to user traffic, those need to be tested. HiLO exposes test-hooks that allow operators and service owners to validate behavior in isolation. A new config is first configured as a test-scenario in HiLO. Frontend service is then directed to tag either test traffic or a fraction of forked production traffic with the test scenario identifier. HiLO applies the candidate states only to this tagged traffic, enabling measurement of OneDCG, latency, and capacity-utilization without affecting real users.

\textbf{Experiment hooks:} After test validation, acceptable states are promoted to experiments (via the existing A/B testing framework~\cite{kohavi:experiments}) using experimentation hooks. Here new states are applied to a fraction of user-traffic (typically 0.5\% over 1 or 3 days and then gradually promoted to 1\% and 5\%) and business metric (e.g., revenue/user, DAU, session-engagement/user) movement is noted. This makes state selection and product configuration a data-driven process.

This methodical approach to identifying states and monitoring them in real-time ensures a high level of predictability in the platform state at any given time.

\subsection{Safety and Operational Properties}
\label{sec:safety}

Some salient properties and design considerations worth understanding are as follows.

\textbf{Eliminating untoward interactions:} A single active state per scenario eliminates ambiguity in product behavior and prevents untoward interactions between independently applied controls.

\textbf{Tuning oscillations:} When load increases, HiLO transitions monotonically toward more restrictive product states. Recovery transitions are damped by a configurable factor, allowing operators to tune the tradeoff between fast recovery and sustained stability.

\textbf{Failure isolation:} Each service enforces the more restrictive of HiLO's assigned level and its own locally determined level. When HiLO is unavailable, services continue to apply their self-determined service levels independently, preserving availability without requiring HiLO's coordination. Once HiLO recovers, coordination at product level is reintroduced without disrupting ongoing service execution.

\section{Evaluation and Results}
\label{sec:eval}
HiLO's performance is evaluated on the following dimensions: operational response to real overload conditions, comparative performance without HiLO, load-factor and capacity improvements during degradation for services and GPU-intensive models, and user-facing impact through A/B experiments and quality metrics.

\subsection{HiLO's Overload Response Evaluated Against Key Metrics}
\label{sec:eval:g-metrics}

\subsubsection{HiLO's Response to Overload}

\begin{figure}[H]
\centering
\includegraphics[width=\columnwidth]{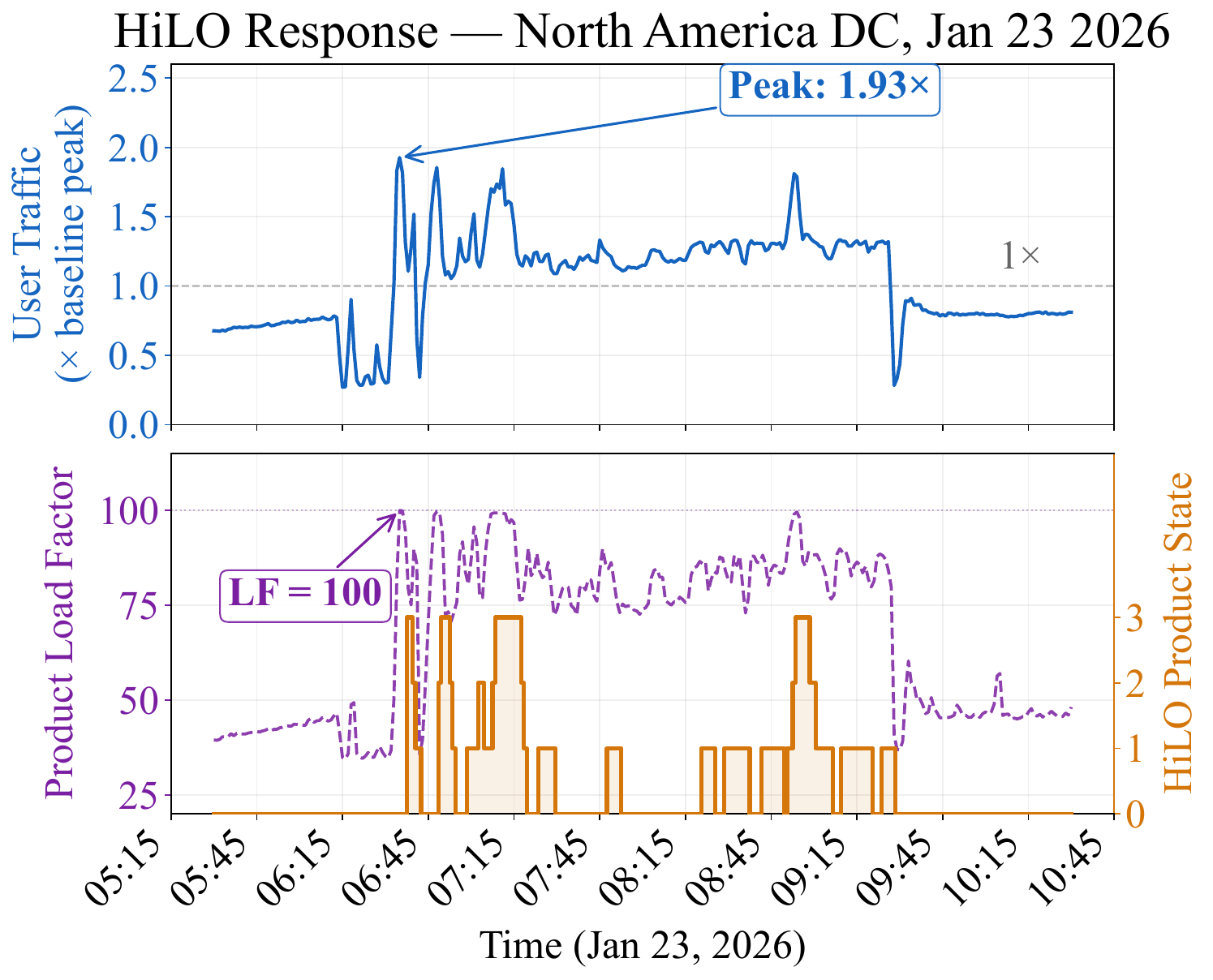}
\caption{HiLO's response to a sustained traffic surge in a North America DC (Jan 23, 2026). Top: user traffic normalized to baseline peak. Bottom: Computed product load-factor and HiLO's product-state selection.}
\label{fig:hilo-response}
\end{figure}

Figure~\ref{fig:hilo-response} presents HiLO's response to one overload event in production. The top chart showcases the increase in traffic (RPS) experienced by a DC in North America. Traffic reached 1.93$\times$ its baseline peak and subsequently stayed in the overload territory for another 3 hours. The bottom chart shows product load-factor as computed and seen by HiLO and its product-state transitions. Quality remained within target limits and stability recovered quickly for this case; discussed further in \S\ref{sec:eval:snapshot}. In a separate model-induced overload (APAC DC, 0.94$\times$ peak, $\sim$6 hours daily over two days), HiLO held OneDCG impact to 0.14 points, versus the 1.11-point drop and severity-2 escalation of the \S\ref{sec:pre-hilo} incident. Note that product-state transitions follow load-factor, not the traffic curve. 1$\times$ baseline peak does not necessarily map to 100\% load-factor (Figure~\ref{fig:hilo-response}) due to headroom and the current query-mix considerations.

\subsubsection{HiLO's One Month Operational Snapshot}
\label{sec:eval:snapshot}

\begin{figure}[H]
\centering
\includegraphics[width=\columnwidth]{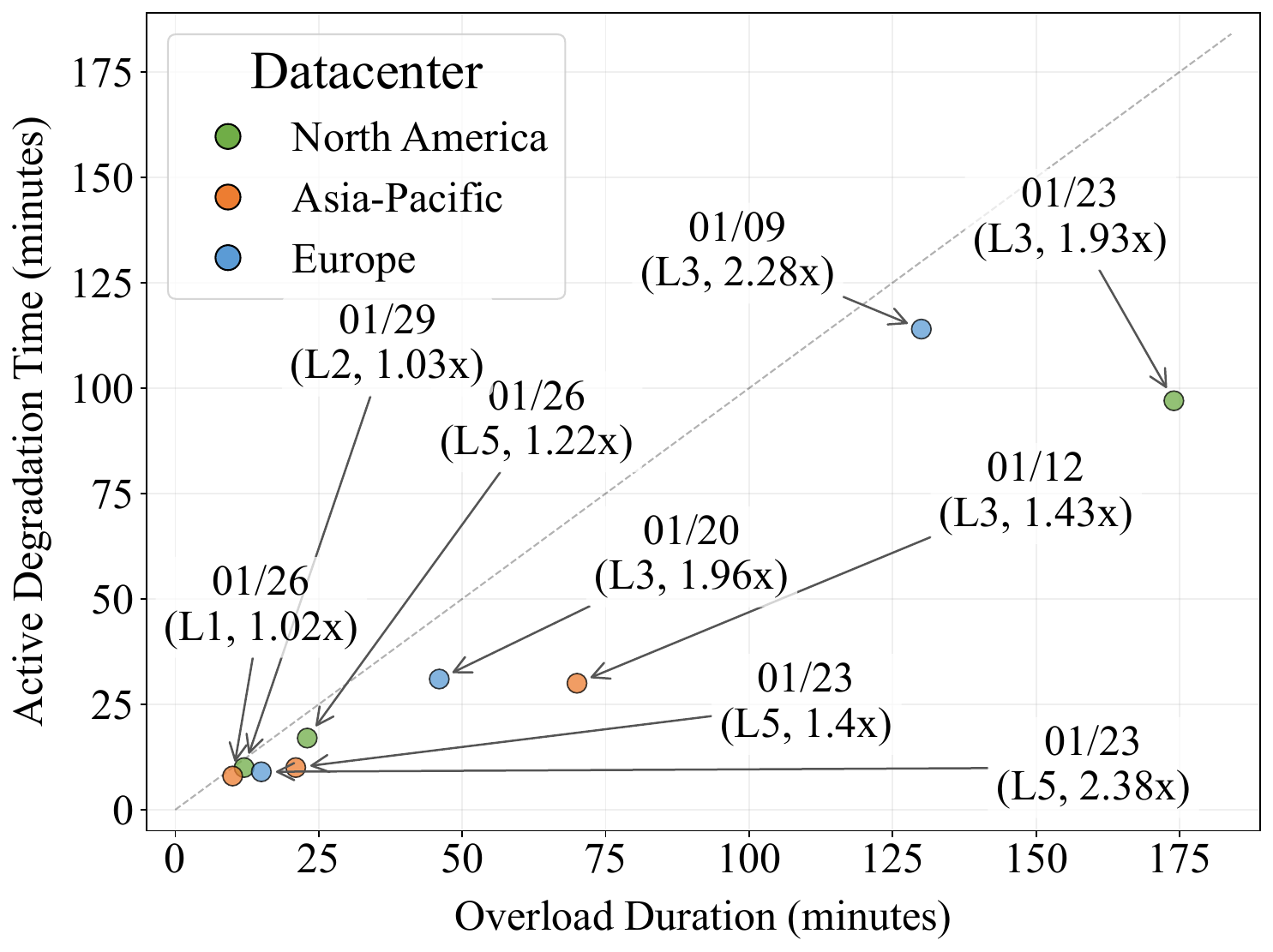}
\caption{HiLO managed degradation during user traffic surge conditions in January 2026. ($L_k$) indicates the max product state reached during degradation.}
\label{fig:scatter-overload}
\end{figure}

During Jan 2026, with HiLO operational across all Bing datacenters (DCs), it actively managed all 9 user-traffic surges that crossed the overload threshold that month, alongside 16 planned capacity tests (a total of 25 events). Figure~\ref{fig:scatter-overload} depicts these 9 user traffic overload events managed by HiLO across three regions varying in intensity (1.02$\times$ to 2.38$\times$ baseline peak) and duration (${\sim}$10~min to $\sim$180~min). The x-axis shows overload duration (first to last minute of user traffic exceeding 1.0$\times$ baseline peak) and the y-axis shows active degradation time (total minutes the product was held at a degraded level, L$_1$ or higher). Baseline peak denotes each datacenter's representative peak traffic level. As noted, some events are fleeting in nature ruling out manual intervention entirely. The figure confirms that active degradation time is always shorter than the overload duration (all points below the diagonal): HiLO degrades only while load-factor exceeds the threshold and recovers promptly. During the rest of the overload, capacity headroom keeps load-factor below the threshold (Figure~\ref{fig:hilo-response}, bottom panel), so the product runs at full quality (L$_0$). 

\medskip
\textbf{Availability:} 
All DCs maintained availability above 99.95\% (target objective) through these events.

\medskip
\textbf{Quality Impact:}

\begin{figure}[H]
\centering
\includegraphics[width=\columnwidth]{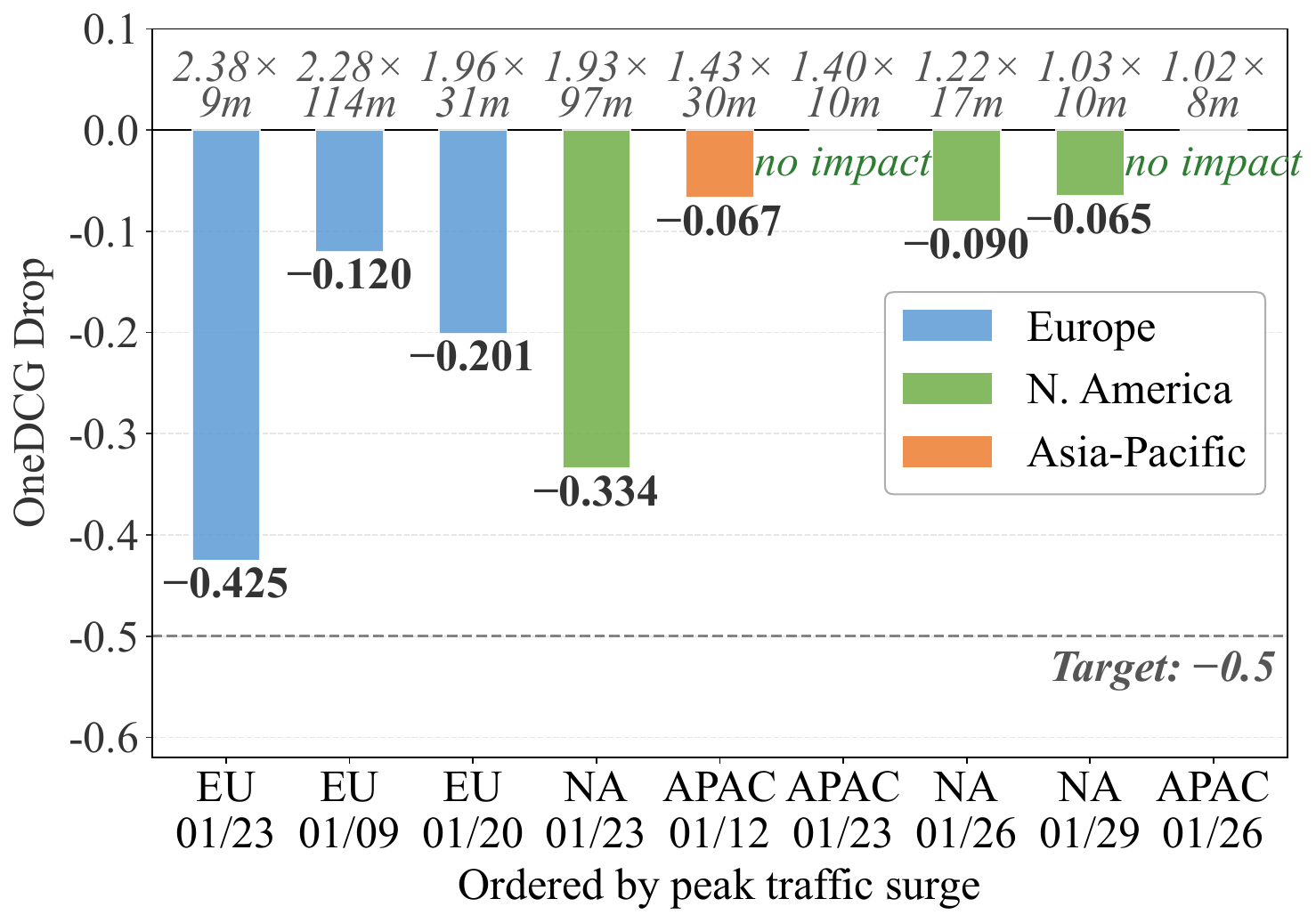}
\caption{OneDCG drop during user traffic surges (Jan 2026), ordered by peak intensity. Annotations show peak traffic ($\times$baseline peak) and degradation duration.}
\label{fig:relevance-drop}
\end{figure}

Figure~\ref{fig:relevance-drop} shows that even the worst-case drop for OneDCG (0.425 points at 2.38$\times$ peak load) remains below the 0.5-point target threshold. No-result rates also remained within normal limits. This demonstrates HiLO's ability to maintain quality under overload.

\begin{figure}[H]
\centering
\includegraphics[width=\columnwidth]{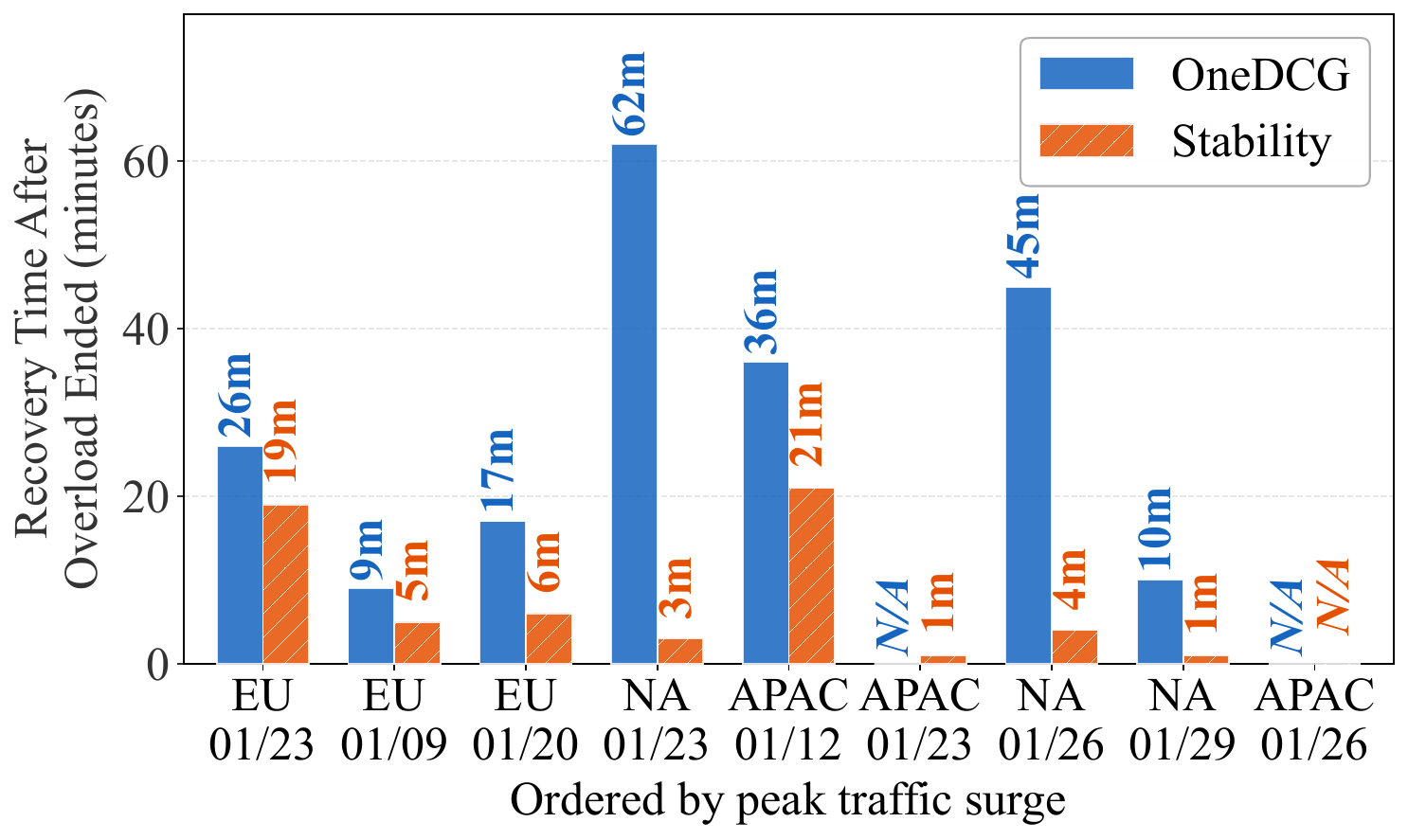}
\caption{Metric recovery after overload ends, measured from the last minute of traffic exceeding 1.0$\times$ baseline peak.}
\label{fig:recovery-combined}
\end{figure}

\medskip
\textbf{Recovery velocity:}
Figure~\ref{fig:recovery-combined} shows metric recovery times. OneDCG recovery is defined as the 5-min rolling avg returning within 0.05 of the pre-event baseline; stability recovery as the first 5-min avg reaching 90\%. Duration from overload end to the first minute of these windows is reported. Metric recovery as noted in the chart (both max and median) confirms HiLO's ability to keep underlying services healthy during overload enabling rapid recovery once load subsides.

\begin{figure}[H]
\centering
\includegraphics[width=\columnwidth]{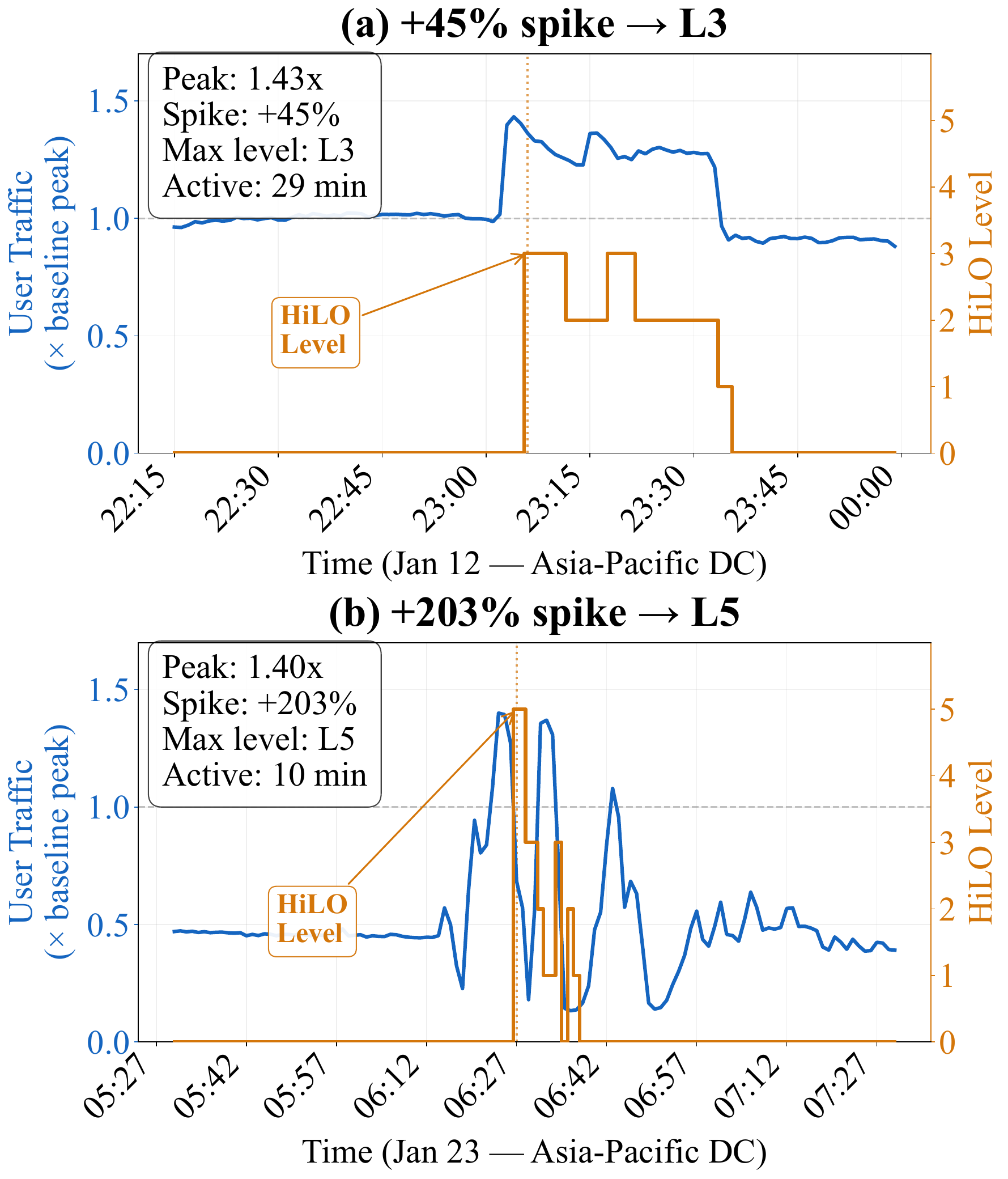}
\caption{HiLO's adaptive response to two different spike profiles.}
\label{fig:burst-vs-gradual}
\end{figure}

\medskip
\textbf{Adaptive response to varying surge profiles:} 
Figure~\ref{fig:burst-vs-gradual} compares two APAC DC events with similar peak traffic but different proportional surges ($+$45\% from 1$\times$ baseline peak vs.\ $+$203\% from 0.5$\times$). HiLO's response is not determined by peak traffic alone; it adapts to load-factor dynamics, producing different escalation depths for the same peak.

\subsubsection{Comparative Performance: Capacity Tests}
\label{sec:eval:capacity}

\begin{figure}[H]
\centering
\includegraphics[width=\columnwidth]{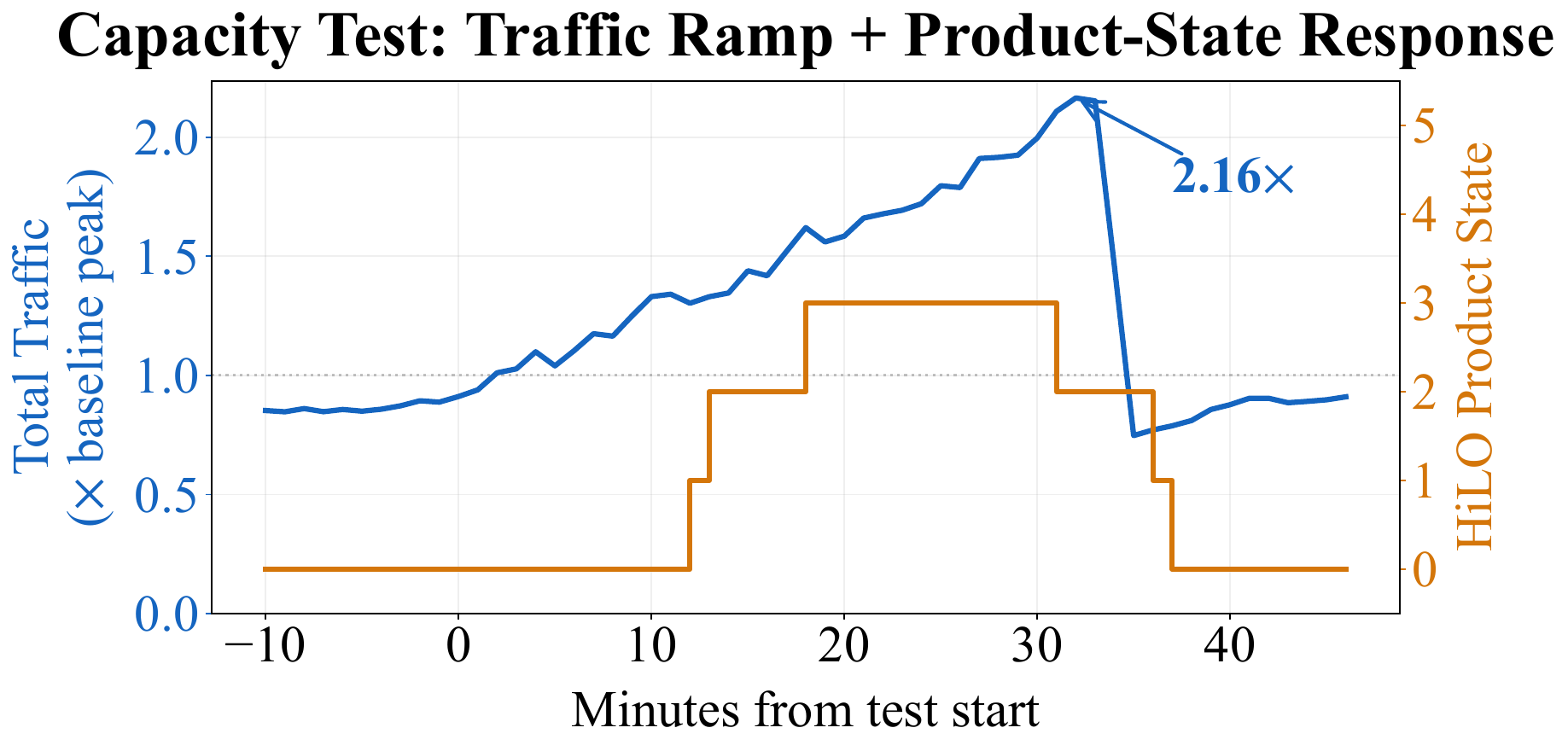}
\caption{Capacity test with HiLO on a Europe DC: normalized traffic (synthetic + organic) and HiLO's response.}
\label{fig:captest-comparison}
\end{figure}

In this section we present capacity test results demonstrating how HiLO improves the system's coordination during degradation. In the baseline (``Without HiLO''), services still degrade under service-local policies (uncoordinated). Bing conducts capacity tests by slowly injecting synthetic traffic to desired level and stopping it. This gradual rampup and HiLO response are shown in Figure~\ref{fig:captest-comparison}. Apparent timing differences between traffic and HiLO's response, including recovery lag after ramp-down and state changes before peak traffic reflect HiLO's decision interval (\S\ref{sec:eval:overhead}), LF signal delays, and degradation actions reclaiming capacity. Without HiLO this product-level behavior was not observable.

Table~\ref{tab:captest-comparison} summarizes the test parameters for capacity tests on two datacenters, without and with HiLO deployment. 

\begin{table}[H]
\centering
\small
\caption{Capacity test comparison: two DCs, without and with HiLO. Peak load vs. OneDCG drop compared.}
\label{tab:captest-comparison}
\setlength{\tabcolsep}{3pt}
\begin{tabular}{@{}l cc cc cc@{}}
\toprule
& \multicolumn{3}{c}{\textbf{Without HiLO}} & \multicolumn{3}{c}{\textbf{With HiLO}} \\
\cmidrule(lr){2-4} \cmidrule(lr){5-7}
& \textbf{Dur.} & \textbf{Peak} & \textbf{OneDCG} & \textbf{Dur.} & \textbf{Peak} & \textbf{OneDCG} \\
& \textbf{(min)} & \textbf{Load} & \textbf{$\Delta$} & \textbf{(min)} & \textbf{Load} & \textbf{$\Delta$} \\
\midrule
APAC & 34 & 2.13$\times$ & -0.05 & 35 & 2.06$\times$ & 0 \\
EU   & 35 & 2.06$\times$ & -0.02 & 36 & 2.16$\times$ & 0 \\
\bottomrule
\end{tabular}
\end{table}

Both tests targeted $\sim$2$\times$ baseline peak load. With HiLO, no OneDCG impact is observed. That said, capacity tests are short-duration events where load ramps gradually; the OneDCG movements discussed in \S\ref{sec:eval:snapshot} provide a more representative picture of impact under production overload.

\begin{figure}[H]
\centering
\includegraphics[width=\columnwidth]{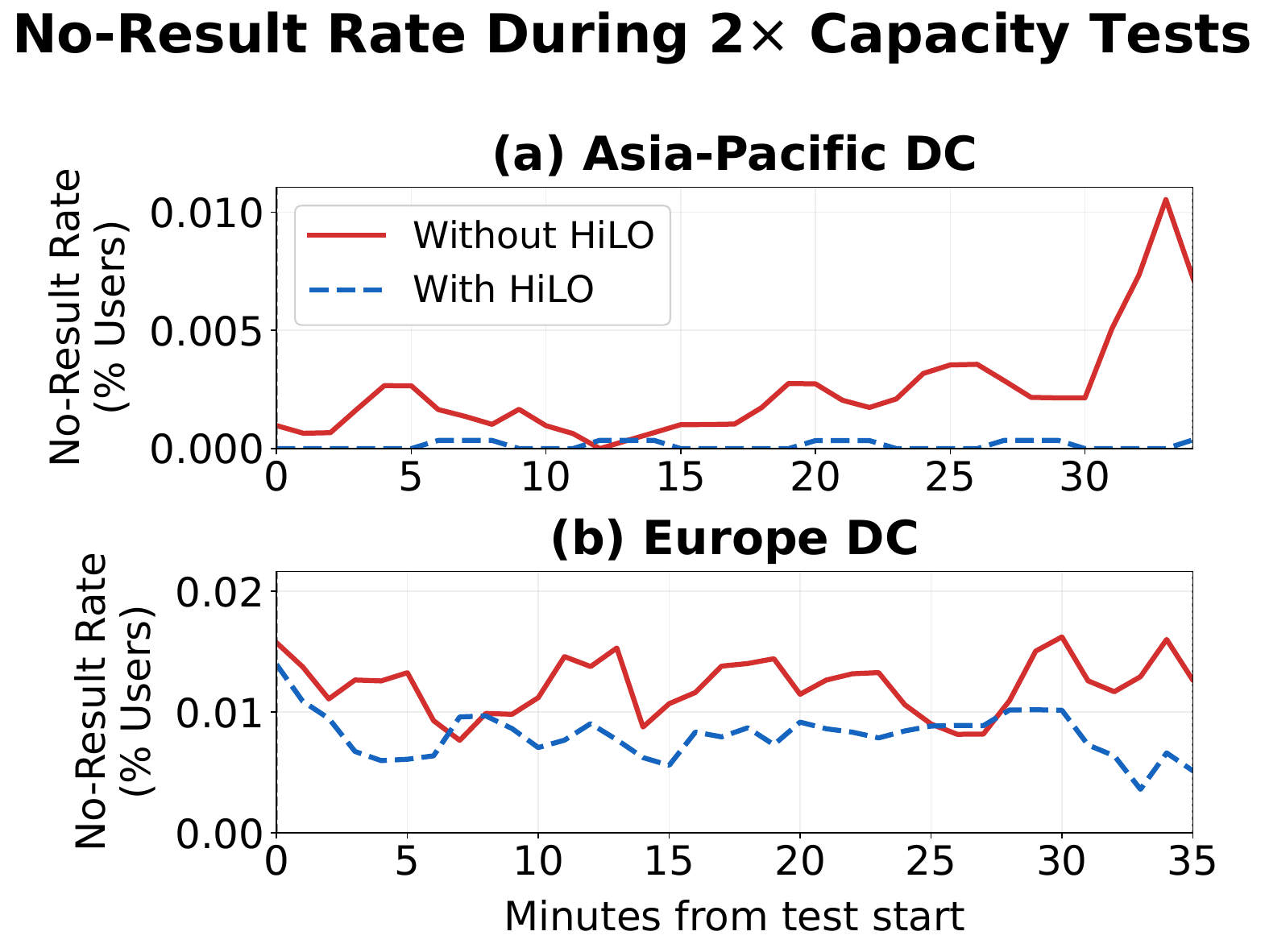}
\caption{No-result error rate (3-min moving avg) with vs.\ without HiLO, across two DCs over the test duration. HiLO's coordinated degradation shows reduced errors. In APAC DC the increase towards the end without HiLO is dramatic.}
\label{fig:captest-noresult}
\end{figure}

Figure~\ref{fig:captest-noresult} shows the difference in No-result error rates with and without HiLO. No-result error rate measures the fraction of queries for which backend services fail to produce results, indicating capacity-induced failures. Without HiLO, these error rates are consistently higher or increase as the test progresses. HiLO now allows for better coordination and lower error rates. With HiLO, operators regularly stress DCs to $\sim$2$\times$ baseline with confidence. This section highlights HiLO's ability to not only provide testability and observability improvements but also to coordinate degradation more effectively under overload conditions.

\subsubsection{30+ Hours, 3 Datacenters and Autonomous Overload Management}
\label{sec:eval:maintenance}

\begin{figure}[H]
\centering
\includegraphics[width=\columnwidth]{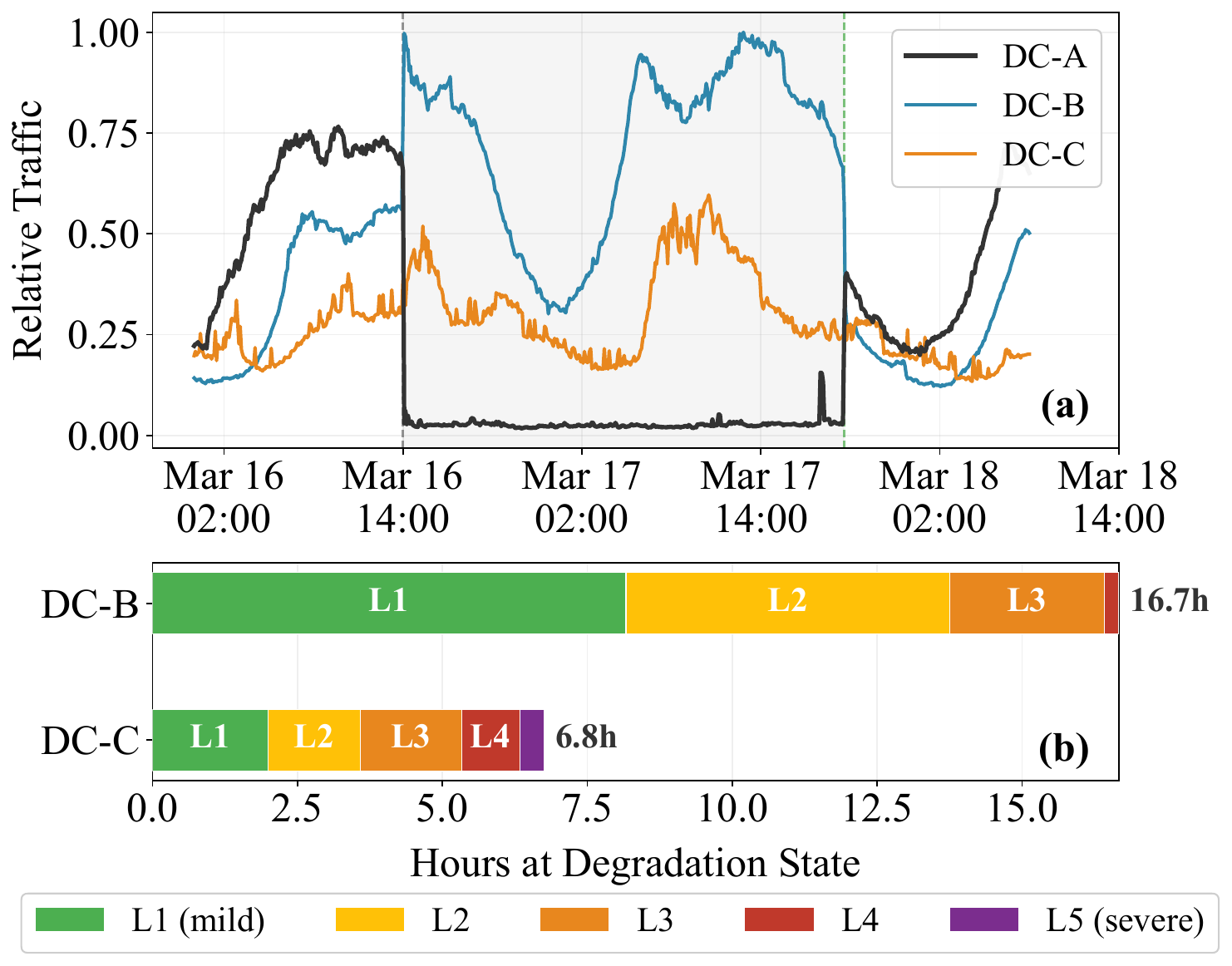}
\caption{(a)~Relative organic traffic across three DCs during planned maintenance. DC-A is taken offline (shaded region), and its load is redistributed to DC-B and DC-C. (b)~HiLO state distribution during the maintenance window.}
\label{fig:maint-traffic-levels}
\end{figure}

Recently HiLO autonomously managed states when a major US DC (DC-A) was taken offline for $\sim$30+ hours for maintenance. Traffic was redistributed by CDN and two other DCs absorbed the load (Figure~\ref{fig:maint-traffic-levels}). Availability remained above target SLO with degradation applied only as necessary. DC-B operated at L$_1$-L$_2$ for 83\% of its 16.7 degradation hours, while DC-C's L$_5$ spikes were brief (0.4 hours).

\begin{figure}[H]
\centering
\includegraphics[width=\columnwidth]{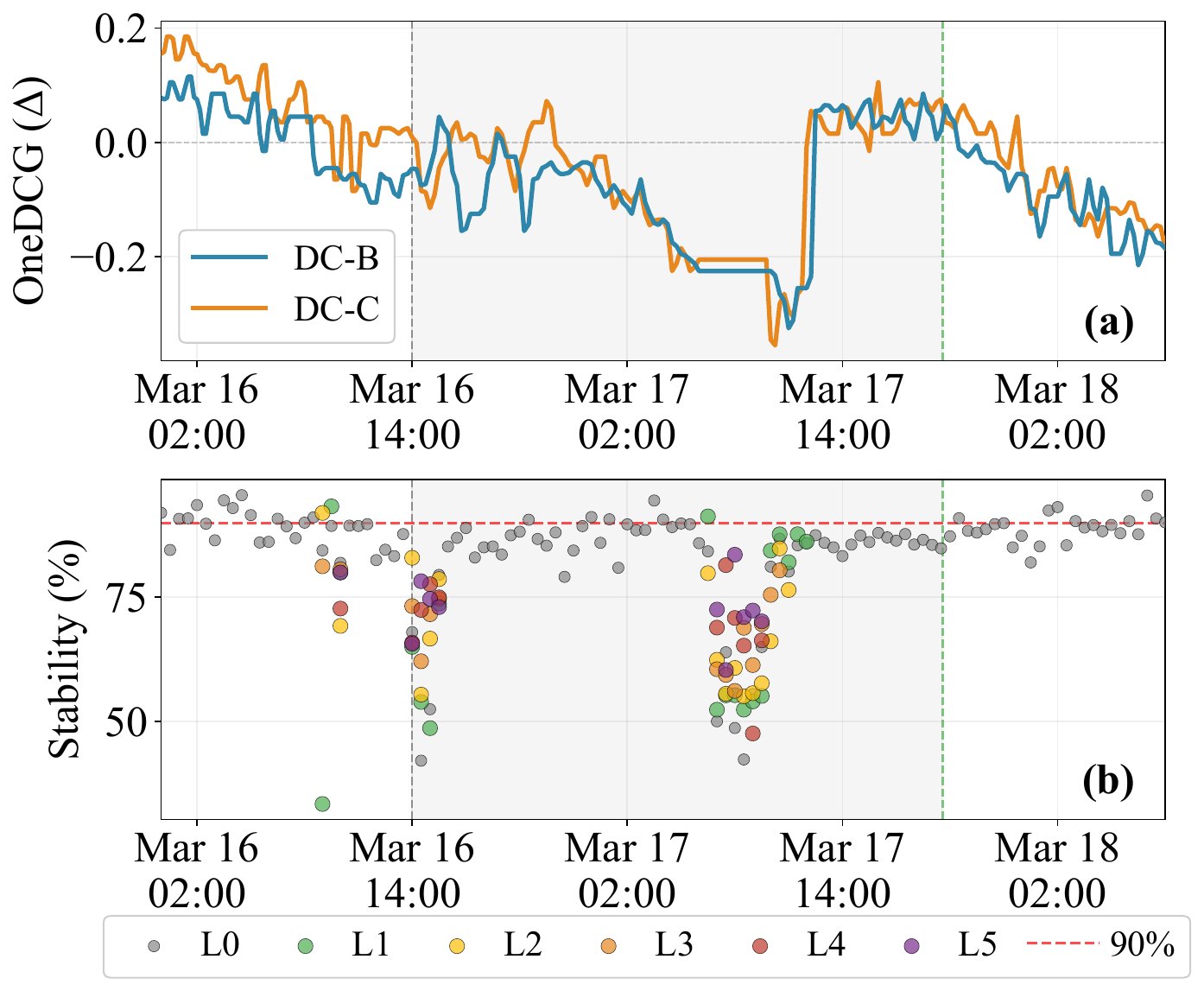}
\caption{(a)~OneDCG values for DC-B and DC-C during this period. DCs remain within normal variance throughout. (b)~ Stability at DC-C, colored by HiLO states, recovers to the baseline immediately after load subsides.}
\label{fig:maint-rel-stability}
\end{figure}

Figure~\ref{fig:maint-rel-stability} demonstrates the ability to track stability by state (previously impossible). OneDCG drop (even during degradation times) was within the target (max drop of 0.38 points). Stability bounced back swiftly once the load normalized. In DC-C, the redirected traffic exposed insufficient capacity for a GPU-intensive model, causing states to oscillate (Figure~\ref{fig:maint-rel-stability}b). Degradation and recovery (\S\ref{sec:safety}) are configurable and reflect business priorities. Under sustained overload, holding a degraded state too long has larger revenue implications (at least for Bing) than those of re-attempting recovery once load-factor falls below the overload threshold. Consequently, in this case, the system oscillated between degradation and recovery until the deficit was resolved (hours later) by provisioning additional model capacity. But unlike \S\ref{sec:pre-hilo} where a similar bottleneck caused a major OneDCG drop and a severity-2 escalation, here HiLO held quality within limits and kept underlying services healthy throughout, requiring no severity-2 interventions. Hence, the key win here is the ability to autonomously maintain the system health and quality under overload, rather than the ability to avoid oscillations entirely during the overload window.

\subsection{Workload Case Studies}
\label{sec:eval:g-casestudies}

\subsubsection{Deep Models and Services Evaluation}
\label{sec:eval:deep-models}

Here we present the levels and corresponding capacity and quality measurements for an expensive deep ranking model and a couple of core services.

\medskip\noindent\textbf{Deep Model: Capacity Gains and Quality Tradeoffs.}
\label{sec:eval:model-evaluation}

Table~\ref{tab:ai-model-eval} provides the degradation modes for a sample ranking model, driven by two independent mechanisms. First, upstream Multi-Queries (MQ) and document reduction progressively reduces inference calls at the model (up to 29\% at L$_5$), creating significant headroom while the model itself operates unchanged. Second, at deeper levels (L$_4$--L$_5$), model layer reduction reduces per-inference latency and compute cost. These metrics are reported independently: call volume reduction acts on demand (fewer calls), while model layer reduction acts on supply (cheaper per call). For comparison, when this model is throttled or skipped entirely, the OneDCG impact far exceeds the Severe threshold ($\gg$0.5), underscoring the value of graduated degradation over binary model disabling. \textit{Note:} model layer reduction is currently validated and pending production deployment. Further degradation levels are being explored simultaneously.

OneDCG is Bing's primary, business-critical quality metric, so service owners carefully manage its drop across levels to keep it within target during degradation. It is not the only signal we track. Broader user-facing online metrics such as content freshness and engagement (\S\ref{sec:eval:experiments}) can degrade more at deeper states even when OneDCG is held.

\begin{table}[H]
\centering
\small
\caption{Ranking model: Two independent capacity mechanisms and their OneDCG impact. Call vol.\ $\Delta$ = inference call change via upstream MQ/doc reduction; Lat.\ $\Delta$ = latency reduction via model layer reduction at constant RPS. OneDCG impact categories: Min.\ ($<$0.1), Mod.\ (0.1--0.25), High (0.25--0.5), Sev.\ ($>$0.5).}
\label{tab:ai-model-eval}
\setlength{\tabcolsep}{4pt}
\begin{tabular}{@{}llrrr@{}}
\toprule
\textbf{State} & \textbf{Action} & \textbf{Call Vol.} & \textbf{Lat. $\Delta$} & \textbf{OneDCG} \\
 & & \textbf{$\Delta$} & \textbf{vs L$_0$} & \textbf{Imp.} \\
\midrule
L$_1$ & MQ + doc reduc. & $\sim$$-$3\% & N/A & Min. \\
L$_2$ & MQ + doc reduc. & $\sim$$-$7\% & N/A & Min. \\
L$_3$ & MQ + doc reduc. & $\sim$$-$13\% & N/A & Min. \\
L$_4$ & Above + layer reduc. & $\sim$$-$22\% & $\sim$$-$7\% & Mod. \\
L$_5$ & Above + layer reduc. & $\sim$$-$29\% & $\sim$$-$12\% & Mod. \\
\bottomrule
\end{tabular}
\end{table}

\medskip\noindent\textbf{Services: Capacity Gains and Quality Tradeoffs.}
\label{sec:eval:services-evaluation}

\begin{table}[H]
\centering
\small
\caption{Balance between capacity gains and OneDCG impact for two services across product states.}
\label{tab:services-eval}
\begin{tabular}{@{}lcrrr@{}}
\toprule
\textbf{Service} & \textbf{State} & \textbf{CPU} & \textbf{Latency $\Delta$} & \textbf{OneDCG} \\
 & & \textbf{$\Delta$ (vs L$_0$)} & \textbf{(vs L$_0$)} & \textbf{Imp.} \\
\midrule
Exec.\ Engine & L$_3$+ & $-$40\% & N/A & Mod. \\
ReRanking  & L$_1$ & $-$5\% & $-$4\% & Min. \\
ReRanking  & L$_2$ & $-$10\% & $-$4\% & Min. \\
ReRanking  & L$_3$ & $-$23\% & $-$14\% & Min. \\
ReRanking  & L$_4$ & $-$33\% & $-$25\% & Mod. \\
ReRanking  & L$_5$ & $-$36\% & $-$25\% & Mod. \\
\bottomrule
\end{tabular}
\end{table}

Table~\ref{tab:services-eval} reports the CPU and latency reductions (capacity gains) and the corresponding OneDCG impact across levels for two services. Exec.\ Engine operates with two levels (full service graph and one degraded level), while ReRanking uses six levels (five of them degraded).

\subsubsection{Case Study: HiLO for Copilot Search}
\label{sec:eval:copilotsearch}

\begin{figure}[H]
\centering
\includegraphics[width=\columnwidth]{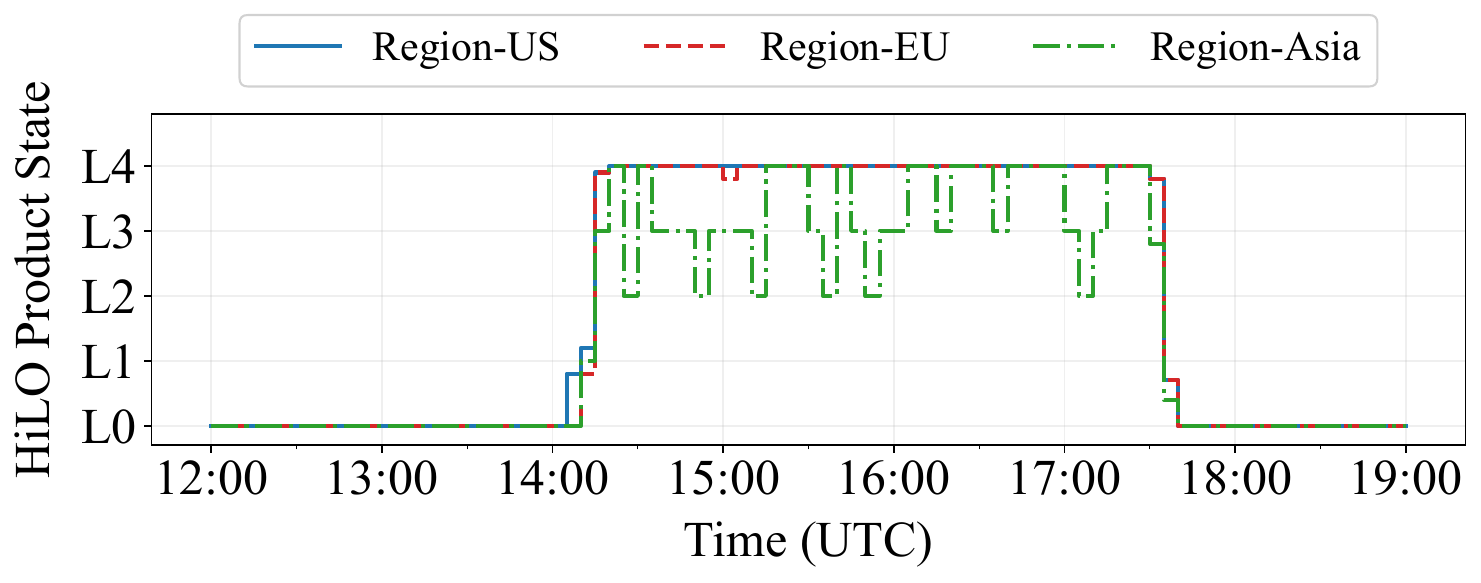}
\caption{HiLO \& Copilot Search during an overload event.}
\label{fig:copilotsearch-levels}
\end{figure}

HiLO applies beyond Web Search and is deployed for Copilot Search, an LLM-centric conversational search experience within Bing. Copilot Search uses 5 product states (L$_0$--L$_4$), with load-factors derived from underlying model services. Figure~\ref{fig:copilotsearch-levels} shows HiLO's product states across three regions during a production overload event. Unlike Web Search, where degradation adjusts backend processing, Copilot Search degrades by progressively disabling product features that trigger model calls: repositioning the Copilot Search tab rightward (L$_1$), disabling additional suggestions (L$_2$), disabling positional features (L$_3$), and hiding the tab entirely (L$_4$). Bing Web Search and Copilot Search represent two distinct execution regimes: fan-out-heavy low-latency services and LLM-centric pipelines with high and variable inference latency. Notably, Copilot Search degradation actions are enforced entirely at the frontend layer (e.g., tab repositioning, feature hiding), whereas Web Search degradation is enforced by backend services. Despite these differences in execution regime and degradation mechanics, HiLO's framework applies unchanged.

\subsection{Operational and User-Facing Impact}
\label{sec:eval:g-impact}

\subsubsection{HiLO Operational Overhead}
\label{sec:eval:overhead}

HiLO operates outside the request path and adds no execution path latency. Table~\ref{tab:overhead} summarizes its operational footprint.

\begin{table}[H]
\centering
\caption{HiLO operational overhead in production.}
\label{tab:overhead}
\begin{tabular}{lc}
\toprule
\textbf{Metric} & \textbf{Value} \\
\midrule
Decision period & 2 minutes \\
HiLO internal latency (P95) & $<$1 ms \\
\# CPU Cores per region & 64 \\
\bottomrule
\end{tabular}
\end{table}

\subsubsection{User-Facing Impact of Product States}
\label{sec:eval:experiments}

HiLO's product states are validated through user-facing A/B experiments before deployment. Each candidate state is activated on a controlled fraction of production traffic, and its impact on key business metrics (e.g., revenue, DAU, session engagement) is measured against baseline (L$_0$). This helps measure the quality tradeoffs at each state and makes state selection a data-driven process.

Table~\ref{tab:ab-experiment} summarizes user-facing impact of product states L$_1$ and L$_2$ vs.\ baseline (L$_0$). Retention is unaffected. Content freshness degrades monotonically, confirming expected tradeoffs under degradation. Distinct Search Query (DSQ) reformulation rate, indicating that users could not find satisfactory results with their initial query, increases consistently across both surfaces and states ($+$0.14\% to $+$3.49\%), with the mobile L$_1$ increase ($+$3.49\%) being statistically significant. This directionally consistent trend confirms a mild but measurable user-facing quality tradeoff under degradation. Tail latency improves on both surfaces, confirming capacity reclamation as an aggregate effect of coordinated degradation across all participating services. Other metrics (e.g., revenue) exhibit similar directional patterns but are omitted here.

\begin{table}[H]
\centering
\caption{User-facing metric impact by product state. Each cell shows the relative change ($\Delta$\%) vs.\ baseline (L$_0$). Metrics marked with $^{*}$ are statistically significant at $p < 0.05$.}
\label{tab:ab-experiment}
\small
\begin{tabular}{@{}p{3.5cm}lrr@{}}
\toprule
\textbf{Metric} & \textbf{Surface} & \textbf{L$_1$ $\Delta$\%} & \textbf{L$_2$ $\Delta$\%} \\
\midrule
\multicolumn{4}{@{}l}{\textit{Retention}} \\
Short-Term DAU          & Both    & $\approx$0\%  & $\approx$0\%  \\
\midrule
\multicolumn{4}{@{}l}{\textit{Content Freshness}} \\
SuperFresh Rate         & Desktop & $-$11.02\%$^{*}$  & $-$15.21\%$^{*}$  \\
SuperFresh Rate         & Mobile  & $-$10.70\%$^{*}$  & $-$14.41\%$^{*}$  \\
\midrule
\multicolumn{4}{@{}l}{\textit{Engagement}} \\
DSQ Reformulation Rate  & Desktop & $+$0.14\%  & $+$0.36\%  \\
DSQ Reformulation Rate  & Mobile  & $+$3.49\%$^{*}$  & $+$1.93\%  \\
\midrule
\multicolumn{4}{@{}l}{\textit{Latency (Server-side)}} \\
WebAnswer Latency P95   & Desktop & $-$1.65\%$^{*}$  & $-$3.30\%$^{*}$  \\
WebAnswer Latency P95   & Mobile  & $-$1.69\%$^{*}$  & $-$3.37\%$^{*}$  \\
\bottomrule
\end{tabular}
\end{table}

\section{Learnings and Limitations}
\label{sec:limitations}

HiLO's design involves deliberate tradeoffs that bound its applicability. This section draws on Bing's experience and guides adoption decisions. Notably, degradation changes are carefully validated and rolled out, reducing the risk of live incidents.

\textbf{Coarse-grained control.} At Bing, HiLO operates over a small number of product states (6 in number), each representing a discrete operating state. This coarse granularity simplifies validation and operational reasoning but limits the system's ability to make fine-grained, per-feature adjustments. In our experience, coarse-grained control is sufficient: finer granularity neither reclaims meaningfully more capacity per step nor accelerates reclamation.

\textbf{Load signals are service-specific.} In our experience, no single indicator works for every service. The right one reflects where a service actually bottlenecks. A network-bound service leans on queue length, since requests pile up when the network is the constraint even while CPU stays low. By contrast, a CPU-bound service, such as ranking, is better captured by CPU utilization. GPU-bound services, such as models and their serving layers, rely on GPU utilization. Some services combine these into a hybrid indicator, such as CPU utilization with queue length, to capture their bottleneck better. The right indicator can also shift as a service's workload evolves. This is precisely where a single load-factor abstraction helps. Each service chooses the indicator, or blend, that matches its own bottleneck, and HiLO adapts automatically because all it requires is a normalized load-factor.

\textbf{Design-time burden.} Defining product states requires each participating service to identify degradation actions, quantify their capacity impact, and compose them into coherent product states. This often takes weeks to months of cross-team coordination. Identifying degradable actions is often non-trivial in services with complex call graphs and dependencies. For example, a service with 90k+ sub-workflows, each using a tiny fraction of CPU, required three months of analysis. Ownership can also be ambiguous: platform teams optimize for load and capacity while feature teams own workflows and user-facing quality, and it is often unclear whether a degradation action belongs in the leaf service or the calling layer. Call-graph analysis reduces manual effort by surfacing degradable actions, but analysis time still grows with service complexity. Full automation of construction remains open, with adaptive, learned control (\S\ref{sec:future}) an active direction. Much of this cost for Bing reflects retrofitting HiLO onto large, pre-existing services as their RPS grew. The framework itself is general and applies to new and existing services alike. A service that treats degradation as a first-class concern from the start onboards with considerably less effort, though some overhead remains for defining and validating degradation options.

\textbf{Staged adoption.} Given this burden, HiLO adoption is better done in stages, and at Bing it proceeded in phases. Services first emitted load-factor signals; coordinated degradation was then applied to the most critical, capacity-constrained backend services and models, and validated end-to-end. Then gradually HiLO coverage was expanded across the stack. Services not yet onboarded continued under their existing service-local degradation until then. At this scale, adoption also requires executive sponsorship and a dedicated central team to drive cross-team coordination.

\textbf{Applicability to decoupled architectures.} A natural question is whether product-level coordination scales to a product built from many small, independently owned services. Managing many decoupled services from a central control plane, out of band, is an established pattern in service meshes. HiLO aggregates per-service status, selects a product state, and distributes coordinated service levels, while each service enforces its assigned level in its own request path. Because HiLO stays off the request path, allows services to fall back to local degradation when unavailable (\S\ref{sec:safety}), and partitions coordination by scenario (\S\ref{sec:further-controls}), the central aggregator is not a choke point. Unlike a service mesh, HiLO coordinates quality-aware degradation and adds no per-request latency (\S\ref{sec:eval:overhead}) as enforcement is by the service itself. For the configuration itself, HiLO coordination is generally applied to the few services on the capacity-constrained critical path, identified at design time (\S\ref{sec:product-levels}). The rest of the services can continue under their own service-local degradation, backed by performance isolation guarantees. In Bing, a given scenario coordinates on the order of a dozen major services and models, so the design cost typically scales with the number of bottleneck services and not the total count.

\textbf{Quality vs. stability tension.} Stability is a choice. Recovery from degradation is governed by a configurable damping factor (\S\ref{sec:safety}). We can achieve a highly stable outcome by enforcing degradation for longer periods. Bing's policy attempts to strike a balance and at times accepts reduced stability during degradation. For example, under 2$\times$ load, aggregate stability drops while OneDCG is preserved (Figure~\ref{fig:maint-rel-stability}). Policy is optimized for business metrics (including revenue) rather than treating stability as a goal in itself.

\textbf{Track degradation itself for operational monitoring.} Effective degradation can obscure underlying issues. In one instance, insufficient model capacity caused daily degradation across multiple DCs for over two weeks. OneDCG and error rates remained within limits, and stability recovered quickly after each incident, masking the underlying issue and making it appear less severe than it was. This led us to treat degradation itself as a first-class signal, tracking its frequency, depth, and duration independently. For example, if a single model causes deeper or longer degradation, then it is escalated as a severity-2 incident.

\textbf{Metric latency and tuning risk.} While availability and OneDCG are measurable in near real-time, business-critical metrics such as revenue and daily active users arrive with multi-hour or next-day latency. A misconfigured product state may appear healthy on fast-loop signals but cause business impact that surfaces only hours later, necessitating both real-time signals and A/B experimentation (\S\ref{sec:eval:experiments}).

\textbf{Load-spreading first, degradation next.} Overload handling happens at two layers. First, increased load in a region is spread globally via CDN routing. Aggregated backend load-factor is also one of the quality signals used for routing at CDN. Graceful degradation at DC level kicks in only if load-spreading proves inadequate. This is why HiLO is scoped per datacenter rather than globally. This mechanism assumes parity across datacenters, without which spreading is infeasible (as features become unavailable). Because the two layers apply independent thresholds and timing to the shared signal, misalignment can still cause feedback. In one observed event, a double traffic spike pointed to such interaction between CDN redistribution and HiLO's response, reinforcing the need for threshold alignment. Bing manages this via appropriate damping factors for failaways and DC-level degradation to prevent rapid oscillations.

\textbf{Continuous investment and optimization.} Periodic validation of states through local tests, capacity tests and A/B experiments is essential. For example, initial A/B experiments revealed that certain mobile surface metrics diverged from expectations, prompting recalibration actions. Based on demonstrated capacity gains, model layer reduction investments are scaled up: exploring deeper degradation levels and expanding coverage to additional models. Also, initially model refresh cycles outpaced degradation validation, prompting us to incorporate degraded-mode design and validation in the regular model development cycle. Similarly, the LFs of services and models outside HiLO's purview are monitored regularly for capacity and quality impact and onboarded to HiLO as needed. Finally, recovery times, particularly stability, vary across DCs and are tracked and studied for optimization opportunities.

\section{Conclusion and Future Work}
\label{sec:future}

Graceful degradation at Bing has evolved over six years. Early approaches prioritized service-local protections and maintained high availability but left product quality as an emergent behavior. HiLO emerged from our experience managing degradation across dependent services and the need to balance availability and quality. HiLO makes quality a controlled outcome by resolving dependencies at design time and autonomously operating over a small lattice of pre-validated product states. Deployed since 2025, HiLO continuously manages overload events across all Bing datacenters (serving over a billion monthly active users~\cite{stats:bing}), bounding quality loss within operational targets across Web Search and LLM pipelines. Our evaluation demonstrates that HiLO preserves availability, bounds quality loss, and produces monotonically ordered user-facing impact validated through A/B experiments, transforming degradation into a predictable and well-controlled process at product level.

Future directions include integrating product-level awareness into cross-region traffic management and exploring learned control policies for HiLO that discover optimal product states using historical load patterns and quality outcomes.

\section*{Acknowledgments}

We thank the teams across Search and AI Platform whose contributions enable coordinated graceful degradation across Bing. This includes frontend, serving platform, relevance, live site engineering, and Bing metrics. We also thank our leadership for their continued support. We finally thank the NSDI review team and our shepherd for their thorough reviews and careful feedback, which materially improved this paper.

\bibliographystyle{plain}
\bibliography{references}

\end{document}